\documentclass[journal]{IEEEtran}
\usepackage{cite}
\ifCLASSINFOpdf
\usepackage[pdftex]{graphicx}
\graphicspath{{../pdf/}{../jpeg/}}
\DeclareGraphicsExtensions{.pdf,.jpeg,.png}
\else
\usepackage[dvips]{graphicx}
\graphicspath{{../eps/}}
\DeclareGraphicsExtensions{.eps}
\fi
\usepackage[cmex10]{amsmath}
\usepackage{array}
\usepackage{mdwmath}
\usepackage{amssymb}
\usepackage{cuted} 
\usepackage{mdwtab}
\usepackage{stfloats}
\usepackage{color}
\usepackage{multirow}
\usepackage{mathrsfs}
\usepackage[tight,footnotesize]{subfigure}
\usepackage{algorithm}  
\usepackage{algorithmic}  
\usepackage{caption}
\usepackage{subcaption}
\usepackage{afterpage}
\usepackage{graphicx}
\usepackage{makecell}

\usepackage{color}

\newtheorem{proposition}{Proposition}

\begin{document}
	\title{Channel-Aware Constellation Design for Digital OTA Computation}
	
	\author{Zeyang~Li, Chen~Chen,~\IEEEmembership{Member,~IEEE}, and~Carlo~Fischione,~\IEEEmembership{Fellow,~IEEE}
		\thanks{This work was funded by the SSF SAICOM project and Digital Futures.
			
			Z. Li, C. Chen and C. Fischione are with the School of Electrical Engineering and Computer Science, and the Digital Futures, KTH Royal Institute of Technology, 114 28 Stockholm, Sweden (e-mail: zeyangl, chch2, carlof@kth.se).}
	}
	
	\markboth{Journal of \LaTeX\ Class Files,~Vol.~14, No.~8, January~2025}%
	{Shell \MakeLowercase{\textit{et al.}}: Bare Demo of IEEEtran.cls for IEEE Journals}
	
	\maketitle
	
	\begin{abstract}
		Over-the-air (OTA) computation has emerged as a promising technique for efficiently aggregating data from massive numbers of wireless devices. OTA computations can be performed by analog or digital communications. Analog OTA systems are often constrained by limited function adaptability and their reliance on analog amplitude modulation. On the other hand, digital OTA systems may face limitations such as high computational complexity and limited adaptability to varying network configurations. To address these challenges, this paper proposes a novel digital OTA computation system with a channel-aware constellation design for demodulation mappers. The proposed system dynamically adjusts the constellation based on the channel conditions of participating nodes, enabling reliable computation of various functions. By incorporating channel randomness into the constellation design, the system prevent overlap of constellation points, reduces computational complexity, and mitigates excessive transmit power consumption under poor channel conditions. Numerical results demonstrate that the system achieves reliable NMSE performance across a range of scenarios, offering valuable insights into the choice of signal processing methods and weighting strategies under varying computation point configurations, node counts, and quantization levels. This work advances the state of digital OTA computation by addressing critical challenges in scalability, transmit power consumption, and function adaptability.
	\end{abstract}
	
	
	\IEEEpeerreviewmaketitle
	
	\section{Introduction}
	\IEEEPARstart{T}{he} next generation of the Internet of Things (IoT) is anticipated to handle and integrate massive amounts of data or computational outputs generated by numerous edge devices \cite{A1}, \cite{A2}. To alleviate the heavy communication burden associated with such large-scale systems, over-the-air (OTA) computation has emerged as an effective and innovative approach. By harnessing the waveform superposition property inherent to the multi-access channel, OTA computation enables fast, efficient, and scalable wireless data aggregation \cite{A3}, \cite{A4}. Unlike traditional digital communication methods, most of the so-far proposed OTA computation methods use an analog modulation scheme, allowing multiple wireless devices to transmit simultaneously over the same time-frequency resources without interference. Notably, the communication resources required by OTA computation do not increase with the number of devices, which fundamentally distinguishes it from conventional digital methods that depend on orthogonal resource allocation \cite{A5}.
	
	While OTA computation holds significant promise for IoT applications, its reliance on analog amplitude modulation imposes notable limitations. Recent efforts have focused on extending OTA computation to digital communication systems. Various methods, including one-bit broadband digital aggregation \cite{A5_2} and frequency-shift keying (FSK) \cite{A5_3}, have been proposed to enable digital OTA computation. These approaches demonstrate potential for specific functions, such as majority voting and sign sum in machine learning tasks. 
	Additionally, a phase-asynchronous OFDM-based version of OBDA has been introduced, incorporating joint channel decoding and aggregation decoders optimized for digital OTA computation \cite{A5_4}. Studies in \cite{A5_3}\cite{A5_5}\cite{A5_6} have addressed non-coherent communication solutions using pulse-position modulation and FSK for both single and multi-cell scenarios. In \cite{A5_8}, the proposed method leverages a balanced numeral system to enable continuous-valued gradient aggregation in a fully digital framework, effectively addressing the limitations of traditional analog OTA Computation approaches, which are susceptible to noise and require precise synchronization. The work presented in \cite{A5_9} introduces an innovative radix-partition-based over-the-air aggregation technique coupled with a low-complexity state estimation scheme for IoT systems operating over wireless fading channels. This method employs pulse amplitude modulation (PAM) to map the decimal representation of input bits, thereby achieving a processing gain.  However, these existing digital methods that we have surveyed above lack generality, being limited to specific functions or modulations. 
	
	To overcome the challenge of limited function adaptability, ChannelComp was introduced, enabling the execution of arbitrary finite functions over the multi-access channel (MAC) through digital modulations \cite{A6}. However, ChannelComp's coding scheme requires considerable computational complexity. To resolve this issue, an enhanced ChannelComp coding scheme was proposed, preserving the advantages of low latency and spectral efficiency associated with both analog OTA computation and ChannelComp \cite{A7}. However, in these works, transmit power consumption of nodes (or mobile devices) can become excessively high, particularly under poor channel conditions, as the transmit power is inversely proportional to the channel quality. Furthermore, the computational complexity is elevated because the constellation used in the demodulation mapping must be designed to prevent overlap between constellation points representing different function values.
	
	Integrating OTA computation with cell-free massive MIMO systems presents an opportunity to leverage the strengths of both paradigms for achieving efficient, low-latency, and energy-conscious wireless communication \cite{A12}, \cite{A13}. On the one hand, cell-free massive MIMO has emerged as a promising technology to provide seamless and uniform wireless coverage while ensuring high spectral and energy efficiency \cite{A8,A9,A10}. Its foundational approach relies on the dense deployment of numerous access points that operate collaboratively to serve all users within the network, effectively eliminating the concept of traditional cell boundaries \cite{A11}. On the other hand, research evaluating the performance of OTA computation within this framework remains relatively sparse. The work in \cite{A12} investigated over-the-air federated learning in scalable Cell-free mMIMO systems but relied on suboptimal maximum ratio combining at the receiver and assumed full-power transmission from wireless devices. The study in \cite{A13} explored OTA computation within Cell-free Massive MIMO systems, explicitly addressing the impact of spatially correlated fading and channel estimation errors. The research proposed optimal methods for designing transmit coefficients and receive combining schemes under varying degrees of cooperation among APs. Nevertheless, these approaches introduce a degree of computational complexity, leading to latency. This latency poses a critical challenge, particularly for applications requiring real-time communication and computation.
	
	In this paper, we propose a digital OTA computation system with channel-aware constellation for demodulation mapper. The major contributions are as follows:
	
	\begin{itemize}
		
		\item A novel digital OTA computation system that utilizes a channel-aware constellation for the demodulation mapper is proposed. The computation point (CP) dynamically adjusts the constellation based on the estimated channel conditions of each node. By incorporating the randomness of channel conditions into the constellation design, overlap between constellation points representing different function values is inherently avoided. As a result, there is no need for a complex constellation design to prevent overlaps, thereby reducing computational complexity and enabling more flexible constellation configurations. Additionally, since the channel effects are inherently considered within the demodulation constellation and do not need to be counteracted, the transmit power no longer needs to scale inversely with channel quality. This eliminates the issue of excessively high transmit power consumption under poor channel conditions.
		
		\item The proposed digital OTA computation system with a channel-aware constellation for the demodulation mapper is capable of supporting a wide range of functions, including sum, product, maximum, and sum of squares. Furthermore, in addition to symmetric functions, the system also accommodates asymmetric functions, as each combination of constellation points from the participating nodes maps uniquely to a corresponding combined constellation point at the CP. This enhancement greatly broadens the scope of potential applications, surpassing the limitations of conventional OTA computation systems that are primarily restricted to simple sum operations.
		
		\item The proposed digital OTA computation system with a channel-aware constellation for the demodulation mapper is applicable to both cellular massive MIMO and cell-free massive MIMO communication systems. For the cell-free OTA computation system, two signal processing approaches are introduced: fully centralized processing and local processing $\&$ centralized voting. In particular, for local processing $\&$ centralized voting scheme, a weighted voting method is proposed, where weights based on channel conditions are assigned to the votes from each CP. Depending on the constraints imposed by the fronthaul overhead between the CPs and the central processing unit (CPU), the most suitable signal processing method can be selected.
		
		\item The numerical results illustrate the normalized mean square error (NMSE) performance of both cellular massive MIMO and cell-free massive MIMO communication systems. Additionally, the NMSE performance is evaluated for various symmetric and asymmetric functions, including sum, product, maximum, and the sum of squares. The results offer valuable insights into the selection of transmit coefficients in cellular massive MIMO systems across different cell ranges, as well as the choice of signal processing techniques and weighted voting methods in cell-free massive MIMO systems under varying numbers of CPs, nodes, and quantization levels.
		
	\end{itemize} 
	
	The rest of this paper is organized as follows: In Section II, we present the system model, detailing the signal model, and the framework for channel-aware constellation design in digital OTA computation. Section III provides an in-depth explanation of the proposed encoding and decoding mechanisms, emphasizing their integration within cellular and cell-free massive MIMO systems. Section IV discusses the performance evaluation of the proposed approach, analyzing NMSE results for various symmetric and asymmetric functions across different configurations, including cellular and cell-free systems. Lastly, we conclude the paper in Section V.
	
	Notation: In this paper, different typographical styles are used to distinguish between scalars, vectors, and matrices. Scalars are represented by italicized letters, vectors by bold lowercase letters, and matrices by bold uppercase letters. The operations $\mathbf{h}^{\mathrm{H}}$ denote the conjugate transpose of a vector $\mathbf{h}$. The $\ell_2$-norm of a vector $\mathbf{h}$ is expressed as $\| \mathbf{h} \|$. The elements in row $i$ and column $j$ of matrix $\mathbf{Y}$, is written as $[\mathbf{Y}]_{i,j}$. The identity matrix of dimension $N \times N$ is denoted by $\mathbf{I}_N$, while the all-ones column vector of size $N$ is represented as $\mathbf{1}_N$. Complex-valued and real-valued matrices of size $M \times N$ belong to the spaces $\mathbb{C}^{M \times N}$ and $\mathbb{R}^{M \times N}$, respectively. The expectation operator is written as $\mathbb{E} \{ \cdot \}$, and the trace of a matrix is given by $\operatorname{tr}(\cdot)$. The notation $\mathcal{N}_c(\mathbf{0}, \mathbf{Y})$ refers to a multivariate circularly symmetric complex Gaussian distribution with zero mean and covariance matrix $\mathbf{Y}$.
	
	\section{System Model}
	In a communication network comprising a CP server and \( K \) nodes, data exchange occurs through a shared MAC. The CP aims to compute a target function \( f(\tilde{x}_1, \tilde{x}_2, \dots, \tilde{x}_K) \), where \( \tilde{x}_k \in \mathbb{F}_q \) represents the input value contributed by node $k$. Each node transmits its value digitally, and the combined transmissions across the MAC enable the CP to perform the necessary computation of \( f \). This setup allows for efficient aggregation of inputs from all nodes to achieve the desired computational goal.
	
	To enable digital transmission, the process typically begins with quantizing each value \( \tilde{x}_k \) into a scalar \( \breve{x}_k \), chosen from a set of \( Q \) discrete levels. Here, \( Q \) corresponds to 2 to the power of the number of quantization bits. The quantized scalar \( \breve{x}_k \) is then transformed into a digitally modulated signal \( x_k \in \{x_{k,1}, x_{k,2}, \dots, x_{k,Q}\} \) through an encoding function \( \mathcal{E}_k(\cdot) \), such that \( x_k = \mathcal{E}_k(\breve{x}_k) \). $x_{k,q}$ denotes one of the possible values that $x_k$ is associated to after the encoding. This encoded signal \( x_k \) is transmitted by node \( k \) across the communication channel.
	
	All nodes transmit concurrently using the same frequency or codes, resulting in the CP server receiving a superimposed signal from all transmitting nodes. The aggregated signal, collected via the MAC during a single time slot, is represented as
	\begin{equation}
		r = \sum_{k=1}^{K} h_k b_k x_k + z \in \mathbb{C},
		\label{received_signal}
	\end{equation}
	where \( r \) is the combined received signal, \( h_k \) denotes the channel coefficient between node \( k \) and the CP server, \( b_k \) is the transmit coefficient of node \( k \), and \( z \) represents additive white Gaussian noise (AWGN) with zero mean and variance \( \delta^2 \).
	
	To estimate the desired function \( f \), the CP server applies an appropriate decoding scheme \( \mathcal{D} \) to the received signal \( r \), i.e.,
	\begin{equation}
		\hat{f} = \mathcal{D}(r),
		\label{EstimatedFunction}
	\end{equation}
	where \( \hat{f} \) is the estimated result of the desired function \( f \).
	
	\section{Channel Aware Constellation for Decoding}
	In this section, we examined the digital over-the-air (OTA) computation using a channel-aware constellation in the simplest scenario, where a single antenna is employed at the CP, i.e., $N_\mathrm{A}=1$. In \cite{A6} and \cite{A7}, for the case of $N_\mathrm{A}=1$, the ideal power control scheme utilized in OTA literature is defined as $b_k = {h_k^*} / {|h_k|^2}$. The desired function value is denoted by
	\begin{equation}
		f = \mathcal{D}(\sum_{k=1}^{K} x_k).
		\label{DesiredFunction0}
	\end{equation} 
	Assuming perfect channel estimation, the received signal at CP is denoted by
	\begin{equation}
		r = \sum_{k=1}^{K} x_k + z.
		\label{received_signal2}
	\end{equation}
	Moreover, the estimated function value is denoted by 
	\begin{equation}
		\hat{f} = \mathcal{D}(\sum_{k=1}^{K} x_k + z).
		\label{EstimatedFunction0}
	\end{equation} 
	The demodulation function $\mathcal{D}(\cdot)$ decodes the received signal according to the combined constellation $\mathbf{s} = [s_1, s_2, \dots, s_M] \in \mathbb{C}^{M \times 1}$, where $s_m=\sum_{k=1}^{K} x_{k, q_{k,m}}$, $q_{k,m} \in [1, 2, \dots, Q]$ indicates which $x_k$ to choose from $\{x_{k,1}, x_{k,2}, \dots, x_{k,Q}\}$ for node $k$, ensuring that the constellation point $s_m$ is received at the CP. $M=Q^K$.
	The goal of designing the constellation $\{x_{k,1}, x_{k,2}, \dots, x_{k,Q}\}$ for modulation mapping is to ensure that distinct function values correspond to unique combined constellation points. This guarantees that the correct function values can be accurately decoded just looking at the received points obtained by the combined constellation $\mathbf{s}$. Mathematically, this requirement can be expressed as $f^{(i)} \neq f^{(j)} \implies s_i \neq s_j$.
	
	Nevertheless, the condition $f^{(i)} \neq f^{(j)} \implies s_i \neq s_j$ may increase the computational complexity of constellation design, particularly when the number of nodes $K$ and the number of possible values $Q$ are large \cite{A6}, \cite{A7}. In addition, the transmit power $b_k = {h_k^*} / {|h_k|^2}$ can occasionally become excessively high. Therefore, we propose channel-aware constellation for decoding to solve these problems.
	
	\subsection{Communication Protocol for Channel-Aware Constellation}
	We assume that the maximum allowable transmit power is $P_\mathrm{t}$. Two types of transmit coefficient are adopted. Specifically, the transmit coefficient of node $k$ is
	\begin{equation} 
		\begin{aligned}
			\text{Type I: } b_k & = \frac{ \sqrt{P_\mathrm{t}} h_k^* }{ |h_k| }; \\
			\text{Type II: } b_k &= \sqrt{P_\mathrm{t}}.
		\end{aligned}
		\label{TransmitCoefficient0}
	\end{equation} 
	In this study, we employ both types of transmit coefficients to evaluate and compare their performance in cellular communication systems. The numerical results present a performance comparison between these two types of transmit coefficients. While both types are applied in cellular communication systems, only Type II is used in cell-free communication systems, as it is the only type applicable in this context.
	
	For Type I of the transmit coefficient, the received signal at the CP is
	\begin{equation}
		r = \sum_{k=1}^{K} h_k b_k x_k + z = \sum_{k=1}^{K} \sqrt{P_\mathrm{t}} |h_k| x_k + z.
		\label{received_signal3}
	\end{equation}
	Furthermore, the estimated function can be denoted by
	\begin{equation}
		\hat{f} = \Tilde{\mathcal{D}} (\sum_{k=1}^{K} \sqrt{P_\mathrm{t}} |h_k| x_k + z).
		\label{EstimatedFunction3}
	\end{equation}
	The demodulation function $\Tilde{\mathcal{D}}$ decodes the received signal $r$ based on a new combined constellation $\Tilde{\mathbf{s}} = [\Tilde{s}_1, \Tilde{s}_2, \dots, \Tilde{s}_M] \in \mathbb{C}^{M \times 1}$, where each element is given by $\Tilde{s}_m = \sum_{k=1}^{K} \sqrt{P_\mathrm{t}} |h_k| x_{k, q_{k,m}}$. The inclusion of $|h_k|$ introduces randomness, which helps prevent overlap between constellation points representing different function values. In other words, the randomness of $|h_k|$ ensures that $f^{(i)} \neq f^{(j)} \implies \Tilde{s}_i \neq \Tilde{s}_j$ is satisfied.
	
	For Type II of the transmit coefficient, the received signal is expressed as
	\begin{equation}
		r = \sum_{k=1}^{K} \sqrt{P_\mathrm{t}} h_k x_k + z.
		\label{received_signal3_2}
	\end{equation}
	Moreover, the estimated function is
	\begin{equation}
		\hat{f} = \Breve{\mathcal{D}} (\sum_{k=1}^{K} \sqrt{P_\mathrm{t}} h_k x_k + z).
		\label{DesiredFunction_2}
	\end{equation}
	The demodulation function $\Breve{\mathcal{D}}$ decodes the received signal $r$ based on a new combined constellation $\Breve{\mathbf{s}} = [\Breve{s}_1, \Breve{s}_2, \dots, \Breve{s}_M] \in \mathbb{C}^{M \times 1}$, where $s_m=\sum_{k=1}^{K} \sqrt{P_\mathrm{t}} h_k x_{k, q_{k,m}}$. In this method, $f^{(i)} \neq f^{(j)} \implies \Breve{s}_i \neq \Breve{s}_j$ is satisfied by introducing the random amplitude and phase of $h_k$.
	
\begin{figure}[t] 
		\centering
		\includegraphics[width=0.48\textwidth]{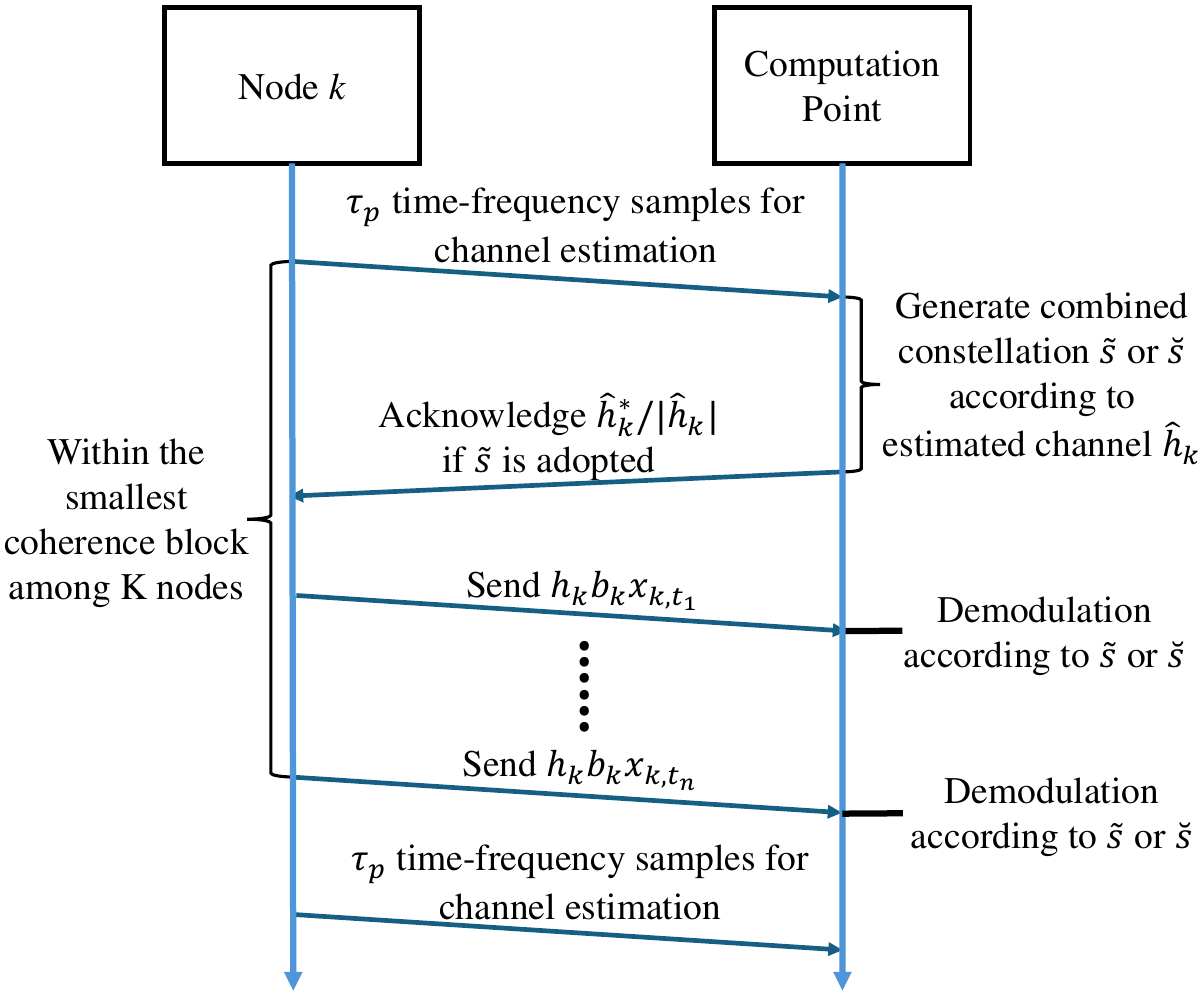} 
		\caption{Communication protocol of the channel-aware constellation framework in OTA computation.}
		\label{ota_framework}
	\end{figure}

	Fig.~\ref{ota_framework} illustrates the communication protocol within a coherence block between node $k$ and the CP adopting channel-aware constellation for demodulation. Initially, \( \tau_p \) time-frequency samples are used for channel estimation. Node $k$ sends pilot signals to allow the CP to estimate the channel ${h}_k$. Based on the estimated channel $\hat{h}_k$, the CP generates a combined constellation $\tilde{\mathbf{s}}$ or $\breve{\mathbf{s}}$. The CP then acknowledges the channel scaling factor $\hat{h}_k^* / |\hat{h}_k|$ if the constellation $\tilde{\mathbf{s}}$ is adopted (if $\breve{\mathbf{s}}$ is adopted, no acknowledgment is needed). Following this, node $k$ transmits its modulated signal $h_k b_k x_{k,t_1}$ and subsequently other signals $h_k b_k x_{k,t_n}$ over time. The CP demodulates the received signals according to the selected constellation $\tilde{\mathbf{s}}$ or $\breve{\mathbf{s}}$. This process takes place within the smallest coherence block among the K nodes, ensuring that the channels of all nodes remain stable during channel estimation, transmission, and demodulation. This stability ensures accurate signal processing and reliable transmission.
	
	For a combined constellation $\Tilde{\mathbf{s}}$ (or $\Breve{\mathbf{s}}$), the Euclidean distance between two constellation points is
	\begin{equation}
		D_{i,j} = |\Tilde{s}_i-\Tilde{s}_j|.
		\label{EuclideanDistanceMat}
	\end{equation}
	The minimum Euclidean distance between the two closest constellation points representing different function values is given by
	\begin{equation}
		d_\mathrm{E} = \text{min} ([D_{i,j}]), i \neq j, f^{(i)} \neq f^{(j)}.
		\label{MiniEuclideanDistance}
	\end{equation}
	\begin{proposition} \label{proposition1}
		Let the constellation design include $|h_k|$ (or $h_k$). Then, $\text{min}([D_{i,j}]) > 0$.
	\end{proposition}
	
	\begin{IEEEproof}
		See appendix A.
	\end{IEEEproof}
	Let $d_\mathrm{R}$ represent the minimum Euclidean distance that the receiver can distinguish in the constellation. For the receiver to detect the Euclidean distance $d_\mathrm{E}$, it must satisfy the condition $d_\mathrm{E} \geq d_\mathrm{R}$. However, $d_\mathrm{E}$ is determined by the channel condition $|h_k|$. Accordingly, if $d_\mathrm{E} < d_\mathrm{R}$, we can amplify the received signal to ensure $d'_\mathrm{E} = A_\mathrm{R} d_\mathrm{E} \geq d_\mathrm{R}$. Therefore, the received signal is amplified by $A_\mathrm{R}$, i.e.,
	
	\begin{equation}
		\begin{aligned}
			r &= A_\mathrm{R} \sum_{k=1}^{K} \sqrt{P_\mathrm{t}} |h_k| x_k + A_\mathrm{R} z, \text{ if } b_k = \frac{ \sqrt{P_\mathrm{t}} h_k^* }{ |h_k| }; \\
			r &= A_\mathrm{R} \sum_{k=1}^{K} \sqrt{P_\mathrm{t}} h_k x_k + A_\mathrm{R} z, \text{ if } b_k = \sqrt{P_\mathrm{t}};
		\end{aligned}
		\label{received_signal3_2_2}
	\end{equation}
	where $A_\mathrm{R}$ is defined as
	\begin{equation}
		A_\mathrm{R} = 
		\begin{cases}
			1, & \text{if } d_\mathrm{E} \geq d_\mathrm{R}; \\
			\frac{d_\mathrm{R}}{d_\mathrm{E}}, & \text{if } d_\mathrm{E} < d_\mathrm{R}.
		\end{cases}
		\label{AmplifyFactor}
	\end{equation}
	
	\begin{figure}[t]
		\centering
		\subfigure[$Q_1=3$ and $Q_2=3$]{
			\raisebox{0pt}[0pt][0pt]{\hspace*{0cm}}{\includegraphics[width=0.7\linewidth]{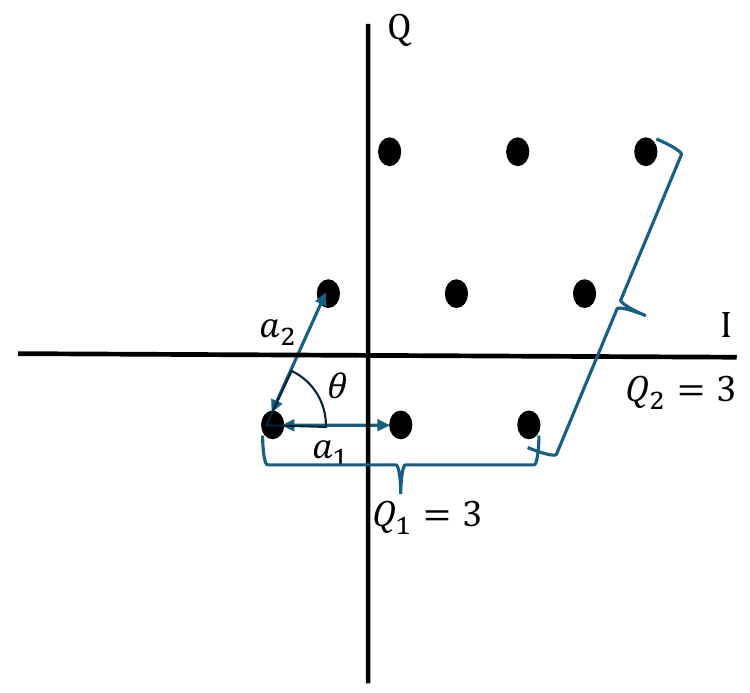}}
		}
		\subfigure[$Q_1=4$ and $Q_2=4$]{
			\raisebox{0pt}[0pt][0pt]{\hspace*{0cm}}\includegraphics[width=0.7\linewidth]{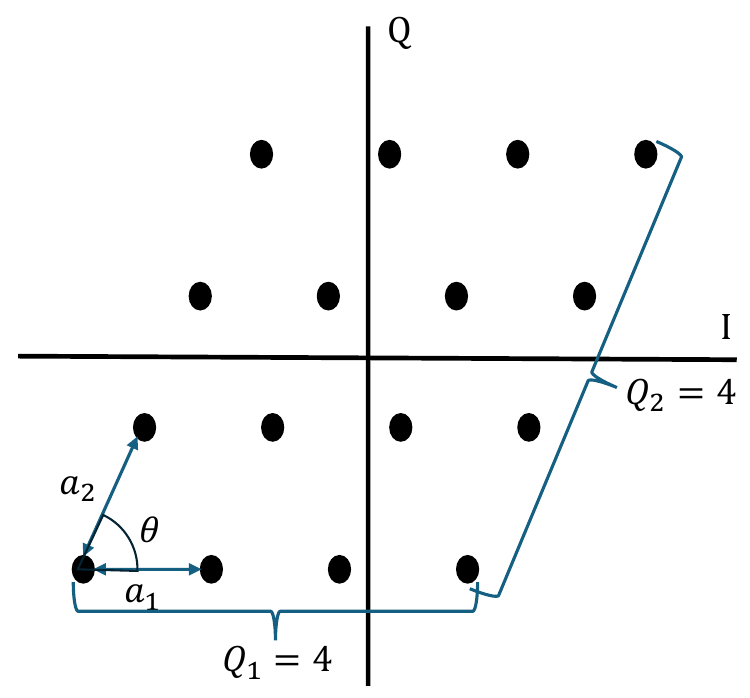}
		}
		\caption{Transmitter Constellation $\mathbf{X}$ Diagrams.}
		\label{TransmitterConstellation}    
	\end{figure}
	
	\begin{figure*}[t]
		\centering
		\subfigure[Constellation $\mathbf{X}_\mathrm{norm}$ for modulation at the transmitters/nodes.]{
			\raisebox{0pt}[0pt][0pt]{\hspace*{0cm}}{\includegraphics[width=0.31\linewidth]{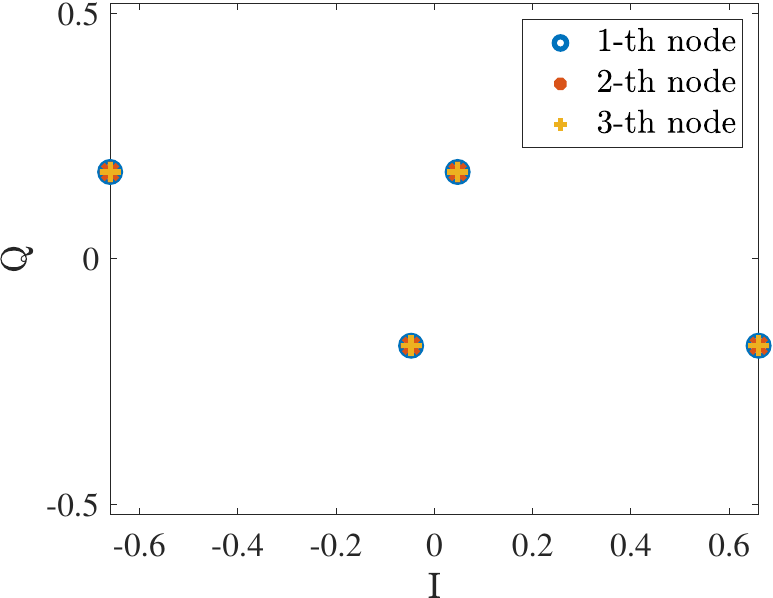}}
		}
		\subfigure[Combined constellation $\mathbf{s}$ for demodulation at the receiver/computation point if $b_k = \frac{h_k^*}{|h_k|^2}$.]{
			\raisebox{0pt}[0pt][0pt]{\hspace*{0cm}}\includegraphics[width=0.32\linewidth]{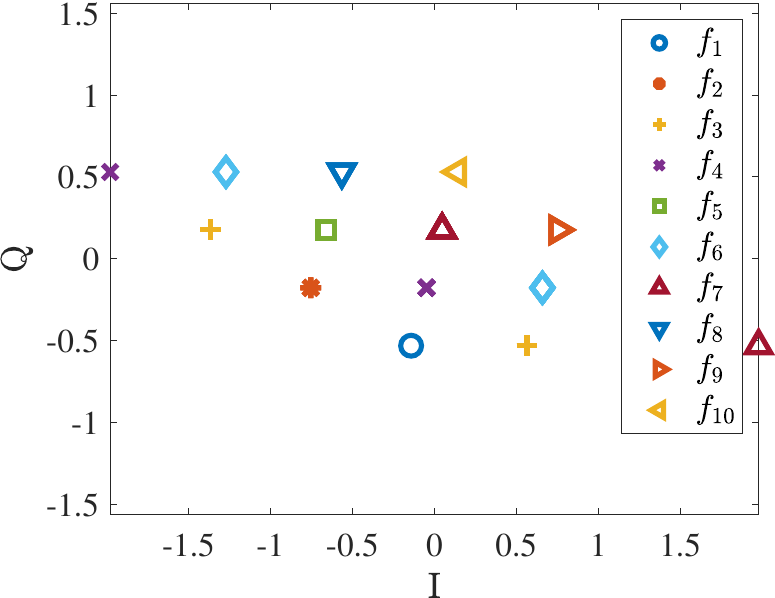}
		}
		\subfigure[Constellation $\sqrt{P_\mathrm{t}} |h_k| \mathbf{X}_\mathrm{norm}$ of node $k$ at the receiver/computation point if $b_k = \frac{ \sqrt{P_\mathrm{t}} h_k^* }{ |h_k| }$.]{
			\raisebox{0pt}[0pt][0pt]{\hspace*{0cm}}{\includegraphics[width=0.31\linewidth]{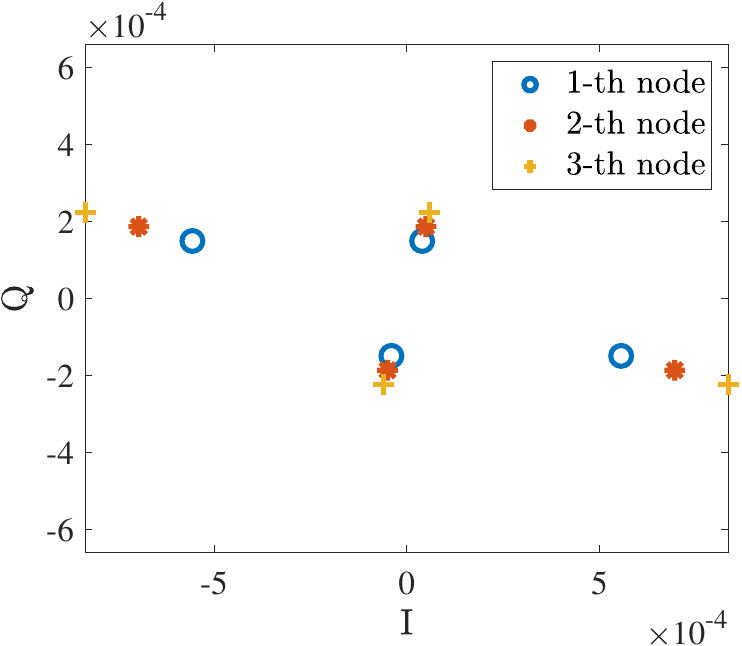}}
		}
		
		\subfigure[Combined constellation $\Tilde{\mathbf{s}}$ for demodulation at the receiver/computation point if $b_k = \frac{ \sqrt{P_\mathrm{t}} h_k^* }{ |h_k| }$.]{
			\raisebox{0pt}[0pt][0pt]{\hspace*{0cm}}\includegraphics[width=0.32\linewidth]{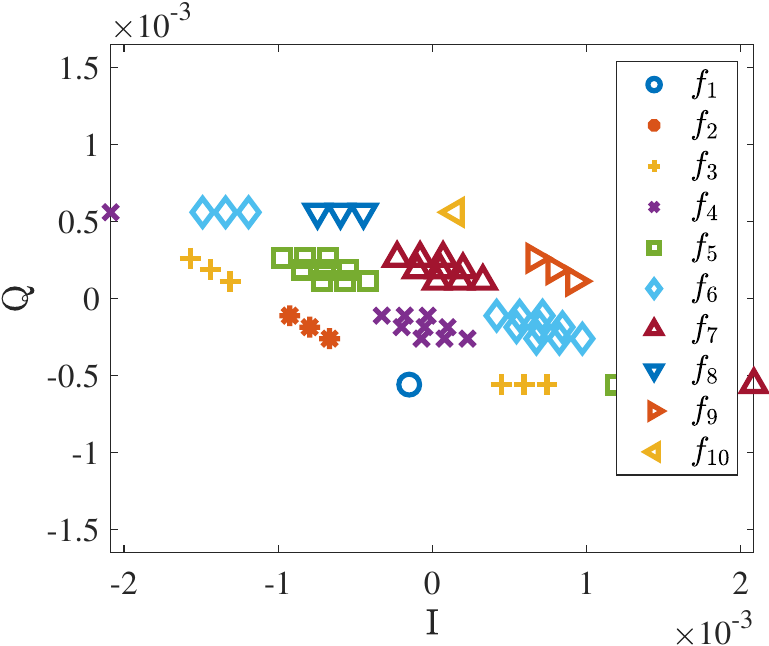}}
		\subfigure[Constellation $\sqrt{P_\mathrm{t}} h_k \mathbf{X}_\mathrm{norm}$ of node $k$ at the receiver/computation point if $b_k = \sqrt{P_\mathrm{t}}$.]{
			\raisebox{0pt}[0pt][0pt]{\hspace*{0cm}}{\includegraphics[width=0.32\linewidth]{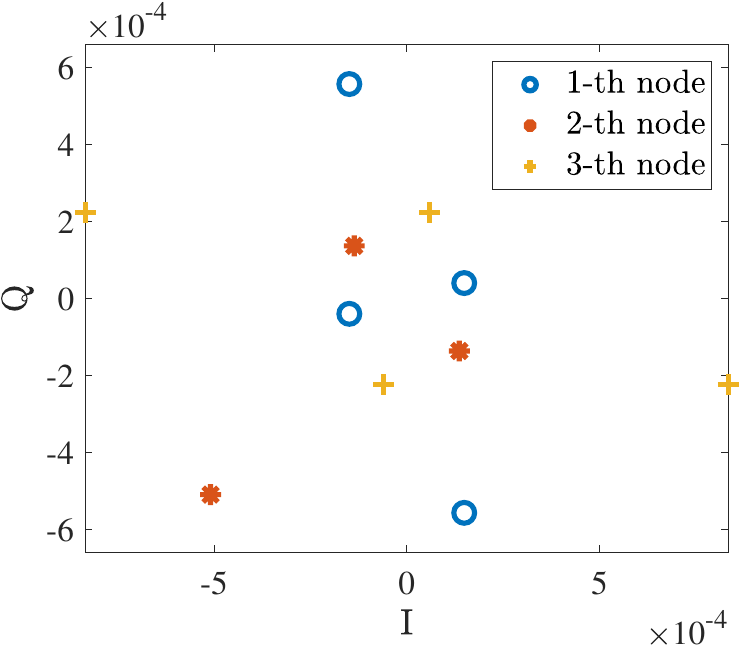}}}
		\subfigure[Combined constellation $\Breve{\mathbf{s}}$ for demodulation at the receiver/computation point if $b_k = \sqrt{P_\mathrm{t}}$.]{
			\raisebox{0pt}[0pt][0pt]{\hspace*{0cm}}\includegraphics[width=0.32\linewidth]{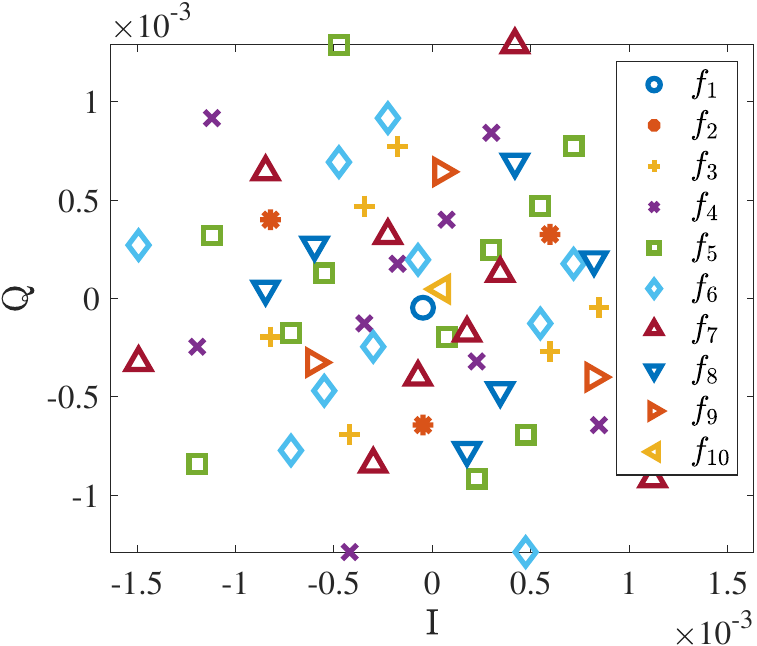}}
		\caption{Constellations for demodulation when different transmission coefficient $b_k$ is adopted when $Q=4$, $Q_1=2$, $Q_2=2$ and $K=3$.}
		\label{TransmitterandReceiverConstellation}    
	\end{figure*}
	
	\subsection{Constellation for Modulation and Demodulation Mapper}
	While the constellation points in $\Tilde{\mathbf{s}}$ and $\Breve{\mathbf{s}}$ do not overlap, a small $d_\mathrm{E}$ can occur, leading to a large $A_\mathrm{R}$. This may exceed the limit $A_\mathrm{RM}$, where $A_\mathrm{R} \leq A_\mathrm{RM}$ must be satisfied if a restriction on $A_\mathrm{R}$ exists at the receiver. To prevent $A_\mathrm{R} > A_\mathrm{RM}$, a specific constellation is introduced for the modulation mapper. On the transmitter side (at the node), the constellation points $[x_{k,1}, x_{k,2}, \dots, x_{k,Q}]$ are selected from $\mathbf{X}_\mathrm{norm}$, which is denoted by 
	\begin{equation}
		\mathbf{X}_\mathrm{norm} = \frac{\mathbf{X}}{\|\mathbf{X}\|_2}.
		\label{NormalizedTransmitConstellation}
	\end{equation}
	An example of $\mathbf{X}_\mathrm{norm}$ is shown in Fig.~\ref{TransmitterandReceiverConstellation} (a). $\mathbf{X}$ is denoted by
	
	\begin{equation}
		\mathbf{X} =
		\begin{cases}
			\mathbf{X}_0 - \frac{a_1 Q_1}{2} - \frac{a_2 Q_2 e^{i \theta}}{2}, & \text{if } Q_1 \text{ and } Q_2 \text{ are odd}, \\
			\mathbf{X}_0 - \frac{a_1 (Q_1 + 1)}{2} - \frac{a_2 (Q_2 + 1) e^{i \theta}}{2}, & \text{otherwise,}
		\end{cases}
		\label{TransmitConstellation1}
	\end{equation}
	where $a_1$, $a_2$, $Q_1$, $Q_2$ ($Q_1 Q_2 \geq Q$) and $\theta$ are pre-decided parameters for the constellation, as illustrated in Fig.~\ref{TransmitterConstellation}. $\mathbf{X}_0$ is denoted by
	\begin{equation}
		\begin{aligned}
			\mathbf{X}_0 = & \{ \text{ } a_1 q_1 + a_2 q_2 \exp{j \theta} \text{ } | \\ & 1 \leq q_1 \leq Q_1, q_1 \in \mathbb{Z}, 1 \leq q_2 \leq Q_2, q_2 \in \mathbb{Z} \}.
		\end{aligned}
		\label{TransmitConstellation2}
	\end{equation}
	At the CP, given the values of $a_1$, $a_2$, $Q_1$, $Q_2$, and $\theta$, the CP can generate the combined constellation $\Tilde{\mathbf{s}}$ or $\breve{\mathbf{s}}$ based on the estimated channel $\hat{h}_k$. Furthermore, the CP can verify whether $A_\mathrm{R} > A_\mathrm{RM}$. If this condition is met, the CP can adjust the values of $a_1$, $a_2$, $Q_1$, $Q_2$, and $\theta$ to ensure that $A_\mathrm{R} \leq A_\mathrm{RM}$.
	
	Fig.~\ref{TransmitterandReceiverConstellation} (b) indicates an example of the combined constellation $\mathbf{s}$ for demodulation at the CP when $b_k = \frac{h_k^*}{|h_k|^2}$. There is an unique constellation point for each function value. However, it is not always guaranteed as $\mathbf{X}_\mathrm{norm}$ changes. Fig.~\ref{TransmitterandReceiverConstellation} (c) shows the constellation $\sqrt{P_t} |h_k| \mathbf{X}_\mathrm{norm}$ for node $k$ at the CP when $b_k = \frac{\sqrt{P_t} h_k^*}{|h_k|}$. Fig.~\ref{TransmitterandReceiverConstellation} (d) illustrates the corresponding combined constellation $\tilde{\mathbf{s}}$ at the CP under the same $b_k$. The constellation points with the same shape and color represent the same function value. Fig.~\ref{TransmitterandReceiverConstellation} (e) presents the constellation $\sqrt{P_t} h_k \mathbf{X}_\mathrm{norm}$ for node $k$ with $b_k = \sqrt{P_t}$. Finally, Fig.~\ref{TransmitterandReceiverConstellation} (f) displays the combined constellation $\breve{\mathbf{s}}$ under $b_k = \sqrt{P_t}$.
	
	\begin{proposition}
		\label{prop:unique_combined_constellation}
		Consider the transmit vectors $\{x_{k,q_{k,m}}\}$ and channel coefficients 
		$\{h_k\}$ (or their magnitudes $\{|h_k|\}$). For each valid combination 
		\begin{equation}
			\sum_{k=1}^{K} \sqrt{P_\mathrm{t}}\,|h_k|\,x_{k,q_{k,m}}
			\quad \text{or} \quad
			\sum_{k=1}^{K} \sqrt{P_\mathrm{t}}\,h_k\,x_{k,q_{k,m}},
			\label{eq:constellation_combination}
		\end{equation}
		there exists a unique mapping to the corresponding combined constellation 
		point $\tilde{s}_m$ (or $\breve{s}_m$) at the CP. Furthermore, beyond symmetric functions, the system also accommodates asymmetric functions due to this one-to-one mapping property.
	\end{proposition}
	
	\begin{IEEEproof}
		See appendix B.
	\end{IEEEproof}
	According to Proposition~\ref{prop:unique_combined_constellation}, the proposed OTA computation systems with channel-aware constellation greatly broaden the scope of potential applications, surpassing the limitations of conventional OTA computation systems that are primarily restricted to symmetric functions.
	
	\subsection{Effect of Channel Estimation Error}
	In this subsection, we introduce channel estimation error, which is inevitable in practical communication system, into the channel-aware constellation system. Consider channel estimation error, Eq. \eqref{received_signal3_2_2} is transformed into
	\begin{equation}
		\begin{aligned}
			r &= A_\mathrm{R} \sum_{k=1}^{K} \frac{ \sqrt{P_\mathrm{t}} h_k \hat{h}_k^* }{ |\hat{h}_k| } x_k + A_\mathrm{R} z, \text{ if } b_k = \frac{ \sqrt{P_\mathrm{t}} \hat{h}_k^* }{ |\hat{h}_k| }; \\
			r &= A_\mathrm{R} \sum_{k=1}^{K} \sqrt{P_\mathrm{t}} h_k x_k + A_\mathrm{R} z, \text{ if } b_k = \sqrt{P_\mathrm{t}},
		\end{aligned}
		\label{received_signal3ChannelError}
	\end{equation}
	where $\hat{h}_k$ is the estimated channel, and the estimation error is \( \tilde{h}_k = h_k - \hat{h}_k \).
	Moreover, for $b_k = \frac{ \sqrt{P_\mathrm{t}} \hat{h}_k^* }{ |\hat{h}_k| }$, the function decoded from the received signal in Eq. \eqref{received_signal3ChannelError} is
	\begin{equation}
		\hat{f} = \Tilde{\mathcal{D}} (A_\mathrm{R} \sum_{k=1}^{K} \frac{ \sqrt{P_\mathrm{t}} h_k \hat{h}_k^* }{ |\hat{h}_k| } x_k + A_\mathrm{R} z).
		\label{DesiredFunctionChannelError}
	\end{equation}
	The Combined constellation for demodulation based on estimated channel is $\Tilde{\mathbf{s}} = [\Tilde{s}_1, \Tilde{s}_2, \dots, \Tilde{s}_M] \in \mathbb{C}^{M \times 1}$, where $\Tilde{s}_m= A_\mathrm{R} \sqrt{P_\mathrm{t}} \sum_{k=1}^{K} |\hat{h}_k| x_{k, q_{k,m}}$.
	On the other hand, for $b_k = \sqrt{P_\mathrm{t}}$, the combined constellation with estimated channel is $\Breve{\mathbf{s}} = [\Breve{s}_1, \Breve{s}_2, \dots, \Breve{s}_M] \in \mathbb{C}^{M \times 1}$, where $\Breve{s}_m=A_\mathrm{R} \sqrt{P_\mathrm{t}} \sum_{k=1}^{K} \hat{h}_k x_{k, q_{k,m}}$.
	
	\section{Extension to Cellular and Cell-Free Massive MIMO systems}
	In this section, the analysis is extended to more complex configurations, including cellular MIMO systems and cell-free massive MIMO systems.
	
	\subsection{Spatially Correlated Rayleigh Fading Channel Model and MMSE Estimation Framework}
	
	Consider a communication system where each CP is equipped with \( N_\mathrm{A} \) antennas, while each node is equipped with a single antenna. The channel between the CP $c$ and node $k$, under the assumption of spatially correlated Rayleigh fading, is modeled as
	\begin{equation}
		\mathbf{h}_{k,c} \sim \mathcal{N}_{\mathbb{C}} \left( \mathbf{0}, \mathbf{Y}_{k,c} \right),
		\label{ChannelCoefficient}
	\end{equation}
	where \( \mathbf{h}_{k,c} \in \mathbb{C}^{N_\mathrm{A}} \) is the channel vector, \( \mathbf{Y}_{k,c} \in \mathbb{C}^{N_\mathrm{A} \times N_\mathrm{A}} \) represents the spatial correlation matrix, and \( N_\mathrm{A} \) is the number of antennas. The local-average channel gain, accounting for path loss and shadowing, is given by \( \beta_{k,c} = \mathrm{Tr}(\mathbf{Y}_{k,c}) / N_\mathrm{A} \). This channel gain, in decibel (dB) scale, is expressed as
	\begin{equation}
		\beta_{k,c}[\mathrm{dB}] = \beta_0 - 10 \alpha \log_{10}\left(\frac{l_{k,c}}{l_0}\right) + S_{k,c},
		\label{PathLoss}
	\end{equation}
	where \( l_0 = 1 \, \mathrm{m} \) denotes the reference distance, \( \beta_0 \) is the large-scale path loss at the reference distance, \( \alpha \) is the path-loss exponent, \( l_{k,c} = \sqrt{l_{{k,c},\mathrm{xy}}^2 + h_\mathrm{CP}^2} \) is the Euclidean distance between the CP $c$ and node $k$, \( l_{{k,c},\mathrm{xy}} \) represents the horizontal distance, \( h_\mathrm{CP} \) is the height difference between the CP and the nodes, and \( S_{k,c} \) is the shadow fading factor.
	
	\subsubsection{Block Fading Channel Model}
	
	The system employs a block fading channel model in which the time-frequency resources are partitioned into coherence blocks of \( \tau_c \) samples. Within a single coherence block, the channel is assumed to be static and frequency-flat, though it may vary across different blocks. For channel estimation, a subset of \( \tau_p \) samples (\( \tau_p \leq K \)) is allocated, with the remaining \( \tau_c - \tau_p \) samples used for signal transmission.
	
	Each node is assigned a pilot signal, \( \boldsymbol{\phi}_k \in \mathbb{C}^{\tau_p} \), selected randomly from a set of \( \tau_p \) mutually orthogonal pilot signals. These signals satisfy the orthogonality condition
	\begin{equation}
		\boldsymbol{\phi}_k^H \boldsymbol{\phi}_i = 0, \quad \forall i \notin \mathcal{P}_k,
	\end{equation}
	where \( \mathcal{P}_k \) denotes the set of nodes sharing the same pilot signal as node \( k \). The norm of each pilot signal is \( |\boldsymbol{\phi}_k|^2 = \tau_p \).
	
	\subsubsection{Channel Estimation}
	
	During the channel estimation phase, the received signal at the CP $c$ is
	\begin{equation}
		\mathbf{R}_c^\mathrm{pilot} = \sum_{k=1}^K \sqrt{p_k} \mathbf{h}_{k,c} \boldsymbol{\phi}_k^H + \mathbf{Z}_c,
		\label{PilotSignal}
	\end{equation}
	where \( p_k \) denotes the pilot transmit power of node \( k \), and \( \mathbf{Z}_c \in \mathbb{C}^{N_\mathrm{A} \times \tau_p} \) is the noise matrix, whose entries are i.i.d. according to \( \mathcal{N}_{\mathbb{C}}(0, \delta^2) \). To estimate the channel, the CP correlates the received signal with the normalized pilot signal \( \boldsymbol{\phi}_k / \sqrt{\tau_p} \), yielding
	\begin{equation}
		\begin{aligned}
			\mathbf{r}_{k,c}^\mathrm{pilot} & = \sum_{i=1}^K \sqrt{\frac{p_i}{\tau_p}} \mathbf{h}_{i,c} \boldsymbol{\phi}_i^H \boldsymbol{\phi}_k + \frac{1}{\sqrt{\tau_p}} \mathbf{Z}_c \boldsymbol{\phi}_k 
			\\ & = \sum_{i \in \mathcal{P}_k} \sqrt{p_i \tau_p} \mathbf{h}_{i,c} + \mathbf{z}_{k,c},
		\end{aligned}
		\label{PilotSignal2}
	\end{equation}
	where \( \mathbf{z}_{k,c} = \frac{1}{\sqrt{\tau_p}} \mathbf{Z}_c \boldsymbol{\phi}_k \sim \mathcal{N}_{\mathbb{C}} \left( \mathbf{0}, \delta^2 \mathbf{I}_{N_\mathrm{A}} \right) \).
	
	The MMSE estimate of \( \mathbf{h}_{k,c} \) is given by~\cite{A13}, \cite{B2}
	\begin{equation}
		\hat{\mathbf{h}}_{k,c} = \sqrt{p_k \tau_p} \mathbf{Y}_{k,c} \boldsymbol{\Xi}_{k,c}^{-1} \mathbf{r}_{k,c}^\mathrm{pilot},
		\label{MMSEChannelEstimate}
	\end{equation}
	where \( \boldsymbol{\Xi}_{k,c} \) is the covariance matrix of \( \mathbf{r}_{k,c}^\mathrm{pilot} \), defined as
	\begin{equation}
		\boldsymbol{\Xi}_{k,c} = \mathbb{E} \left\{ \mathbf{r}_{k,c}^\mathrm{pilot} \left( \mathbf{r}_{k,c}^\mathrm{pilot} \right)^H \right\} = \sum_{i \in \mathcal{P}_k} p_i \tau_p \mathbf{Y}_{i,c} + \delta^2 \mathbf{I}_{N_\mathrm{A}}.
		\label{MMSEChannelEstimate2}
	\end{equation}
	Furthermore, the estimation error is \( \tilde{\mathbf{h}}_{k,c} = \mathbf{h}_{k,c} - \hat{\mathbf{h}}_{k,c} \).
	
	\subsection{Effect of Multiple Antennas at the CP}
	In this subsection, multiple antennas at the CP for channel-aware constellation for demodulation in digital OTA computation are investigated. Consider $N_\mathrm{A} \geq 1$, the received signal at the CP is
	\begin{equation}
		r_\mathrm{mul} = A_\mathrm{R} \sum_{k=1}^{K} \sum_{n_\mathrm{A}=1}^{N_\mathrm{A}} h_{k,n_\mathrm{A}} b_k x_k + A_\mathrm{R} \sum_{n_\mathrm{A}=1}^{N_\mathrm{A}} z_{n_\mathrm{A}},
		\label{received_signal3_3}
	\end{equation}
	where $h_{k,n_\mathrm{A}}$ is the channel coefficient between node $k$ and the $n_\mathrm{A}$-th antennas of the CP, and $z_{n_\mathrm{A}} \sim \mathcal{N}_{\mathbb{C}} \left( \mathbf{0}, \delta^2 \right)$. The two types of transmit coefficient $b_k$ are
	\begin{equation} 
		\begin{aligned}
			\text{Type I: } b_k & = \frac{ \sqrt{P_\mathrm{t}} \sum_{n_\mathrm{A}=1}^{N_\mathrm{A}} \hat{h}_{k,n_\mathrm{A}}^* }{ | \sum_{n_\mathrm{A}=1}^{N_\mathrm{A}} \hat{h}_{k,n_\mathrm{A}}| }; \\
			\text{Type II: } b_k &= \sqrt{P_\mathrm{t}}.
		\end{aligned}
		\label{TransmitCoefficientChannelErrorMultipleAntennas}
	\end{equation} 
	The corresponding combined constellation is: $\Tilde{s}_m= A_\mathrm{R} \sqrt{P_\mathrm{t}} \sum_{k=1}^{K} | \sum_{n_\mathrm{A}=1}^{N_\mathrm{A}} \hat{h}_{k,n_\mathrm{A}}| x_{k, q_{k,m}}$ for Type I; $\Breve{s}_m= A_\mathrm{R} \sqrt{P_\mathrm{t}} \sum_{k=1}^{K}  \sum_{n_\mathrm{A}=1}^{N_\mathrm{A}} \hat{h}_{k,n_\mathrm{A}} x_{k, q_{k,m}}$ for Type II.
	\subsection{Cell-Free Communication System}
	In this subsection, the channel-aware constellation design for demodulation in digital OTA computation within a cell-free communication system is analyzed. In this system, $b_k= \sqrt{P_\mathrm{t}}$, and $C$ receivers/CPs are participated. Two distinct approaches for processing the signals received from the participating CPs are adopted: \\
	\textbf{Fully Centralized Processing:}
	
	For fully centralized processing (FCP), all the received signals from the CPs are forwarded to the central CPU via the fronthaul and processed there.
	The received signal in $c$-th CP is expressed as
	\begin{equation}
		r_c^\mathrm{cel0} = \sqrt{P_\mathrm{t}} \sum_{k=1}^{K} \sum_{n_\mathrm{A}=1}^{N_\mathrm{A}} h_{k,n_\mathrm{A}}^c x_k + \sum_{n_\mathrm{A}=1}^{N_\mathrm{A}} z_{n_\mathrm{A},c}.
		\label{received_signal3_5_2}
	\end{equation} 
	where $h_{k,n_\mathrm{A}}^c$ is the channel coefficient between node $k$ and the $n_\mathrm{A}$-th antennas of the $c$-th CP, and $z_{n_\mathrm{A},c} \sim \mathcal{N}_{\mathbb{C}} \left( \mathbf{0}, \delta^2 \right)$. Furthermore, the received signal at the central CPU is denoted by
	\begin{equation}
		\begin{aligned}
			r_\mathrm{cel} & = A_\mathrm{R} \sum_{c=1}^{C} r_c^\mathrm{cel0} \\ & = A_\mathrm{R} \sqrt{P_\mathrm{t}} \sum_{k=1}^{K} \sum_{c=1}^{C} \sum_{n_\mathrm{A}=1}^{N_\mathrm{A}} h_{k,n_\mathrm{A}}^c x_k + A_\mathrm{R} \sum_{c=1}^{C} \sum_{n_\mathrm{A}=1}^{N_\mathrm{A}} z_{n_\mathrm{A},c}.
		\end{aligned}
		\label{received_signal3_5}
	\end{equation}
	Moreover, $r_\mathrm{cel}$ is decoded, i.e., $\hat{f} = \breve{\mathcal{D}}(r_\mathrm{cel})$. FCP results in a significant burden on the fronthaul overhead, especially when the number of nodes is large, i.e., $K\gg1$. Specifically, the $c$-th CP requires to transmit \{ $\hat{h}_{k,n_\mathrm{A}}^c |$  $k \in [1,K]$, $k \in \mathbb{Z}$, $n_\mathrm{A} \in [1,N_\mathrm{A}]$, $n_\mathrm{A} \in \mathbb{Z}$ \} and $r_c^\mathrm{cel0}$ to the central CPU. To address this issue, the following approach is proposed as an alternative solution to alleviate the fronthaul overhead in cell-free communication systems.
	\\
	\textbf{Local Processing $\&$ Centralized Voting:}
	
	For the approach of local processing $\&$ centralized voting (LPCV), each CP decodes the received signal separately. The received signal in $c$-th CP is expressed as
	\begin{equation}
		r_c^\mathrm{cel} = A_{\mathrm{R},c} \sqrt{P_\mathrm{t}} \sum_{k=1}^{K} \sum_{n_\mathrm{A}=1}^{N_\mathrm{A}} h_{k,n_\mathrm{A}}^c x_k + A_{\mathrm{R},c} \sum_{n_\mathrm{A}=1}^{N_\mathrm{A}} z_{n_\mathrm{A},c},
		\label{received_signal3_6}
	\end{equation}
	where $A_{\mathrm{R},c}$ is the amplification factor of the received signal at the $c$-th CP. Moreover, $r_c^\mathrm{cel}$ is decoded, i.e., $\hat{f}_c = \breve{\mathcal{D}}(r_c^\mathrm{cel})$.
	The estimated function value $\hat{f}_c$ is computed at the $c$-th CP and transmitted to the central CPU, where a voting process is conducted. The final estimated function value $f_{\text{mode}}$ is determined as the majority value of $f$. The vote process is denoted by
	\begin{equation}
		f_{\text{mode}} = \underset{f \in \{f_1, f_2, \dots, f_W\}}{\arg\max} \, \sum_{c=1}^C \omega_c \cdot \mathbb{I}(\hat{f}_c = f),
		\label{EstimatedFunction2}
	\end{equation}
	Here, $W$ represents the number of possible values that $f$ can be, and $\omega_c$ denotes the weight assigned to the vote of the $c$-th CP, reflecting its importance. When $\omega_c = 1$, it indicates that each CP has an equal vote in determining the value of $f_{\text{mode}}$, and each node only needs to transmit $\hat{f}_c$ to the central CPU.
	
	$\omega_c$ can be determined based on the channel conditions of each CP. Therefore, we propose the other $\omega_c$ assignment option 
	\begin{equation}
		\omega_c = \frac{S_c}{ \sum_{j=1}^{C} S_j }.
		\label{AssignmentOption1}
	\end{equation}
	where $S_c$ is expressed as Eq. \eqref{AssignmentOption0}.
	\begin{figure*}
		\begin{equation}
			S_c = \frac{|\sum_{k=1}^{K} \sum_{n_\mathrm{A}=1}^{N_\mathrm{A}} \hat{h}_{k,n_\mathrm{A}}^c|^2}{
				1+ \text{max} \{ | \sum_{n_\mathrm{A}=1}^{N_\mathrm{A}} \hat{h}_{k,n_\mathrm{A}}^c |^2 \big| k \in [1,K] \} - \text{min} \{ | \sum_{n_\mathrm{A}=1}^{N_\mathrm{A}} \hat{h}_{k,n_\mathrm{A}}^c |^2 \big| k \in [1,K] \} }.
			\label{AssignmentOption0}
		\end{equation}
	\end{figure*}
	According to Eq. \eqref{AssignmentOption0}, \(\omega_c\) increases when the total power of the sum of channels from all nodes to the \(k\)-th CP, i.e., $|\sum_{k=1}^{K} \sum_{n_\mathrm{A}=1}^{N_\mathrm{A}} \hat{h}_{k,n_\mathrm{A}}^c|^2$, rises, and the channel power difference between nodes, i.e., $\text{max} \{ | \sum_{n_\mathrm{A}=1}^{N_\mathrm{A}} \hat{h}_{k,n_\mathrm{A}}^c |^2 \big| k \in [1,K] \} - \text{min} \{ | \sum_{n_\mathrm{A}=1}^{N_\mathrm{A}} \hat{h}_{k,n_\mathrm{A}}^c |^2 \big| k \in [1,K] \}$, drops. In other words, Eq. \eqref{AssignmentOption0} rewards having a strong overall sum of channel conditions while penalizing large variations of channel conditions among nodes. For the weight assignment in Eq. \eqref{AssignmentOption1}, each CP needs to transmit $\hat{f}_c$ and $S_c$ to the central CPU.
	
	\section{Numerical Results}
	In this section, we present the NMSE metric performance, which is defined as
	\begin{equation} 
		\text{NMSE} := \frac{\sum_{j=1}^{N_\mathrm{s}} \left| f_j - \hat{f}_j \right|^2}{N_s \left| f_\mathrm{max} - f_\mathrm{min} \right|^2},
		\label{NMSEDefination}
	\end{equation} 
	where $N_\mathrm{s}$ denotes the number of Monte Carlo trials, $f_j$ denotes the value of the desired function we wish to compute in the $j$-th trial, $\hat{f}_j$ is the estimated value of $f_j$, $f_\mathrm{max}$ and $f_\mathrm{min}$ are the maximal and minimal value that the function $f$ can be. The spatial correlation matrix \( \mathbf{Y}_k \) is simulated as
	\begin{equation}
		[\mathbf{Y}_{k}]_{i,j} 
		= \beta_k \rho^{\lvert i-j \rvert},
		\quad i,j = 1,2,\dots,M,
	\end{equation}
	where $\rho=0.8$.
	The system parameters are shown in Table \ref{simulationparameters} if not mentioned otherwise.
	
	\begin{table}[t]
		\begin{center}
			\caption{Simulation Parameters}
			\scalebox{1}{
				\begin{tabular}{|c|c|c|c|l}
					\cline{1-4}
					parameter         & value & parameter & value   &  \\ \cline{1-4}
					$a_1$ & 1 & $a_2$ & 1 &  \\ \cline{1-4}
					$l_0$    & 1  & $\beta_0$   & -30.5 dB  &  \\ \cline{1-4}
					$\alpha$    & 3.67   & $S_k$     & $\mathcal{N}(0, 4^2)$ &  \\ \cline{1-4}
					$h_\mathrm{CP}$    & 20 m   & $P_\mathrm{t}$       & 20 dBm   &  \\ \cline{1-4}
					$\delta^2$          & -96 dBm   & $\theta$       & $\pi/3$      &  \\ \cline{1-4}
					$\rho$          & 0.8   & $\tau_p$       & $K$      &  \\ \cline{1-4}
			\end{tabular}}
			\label{simulationparameters}
		\end{center}
	\end{table}
	
	\begin{figure*}[t]
		\centering
		\subfigure[Symmetric function: sum and product.]{
			\raisebox{0pt}[0pt][0pt]{\hspace*{0cm}}{\includegraphics[width=0.45\linewidth]{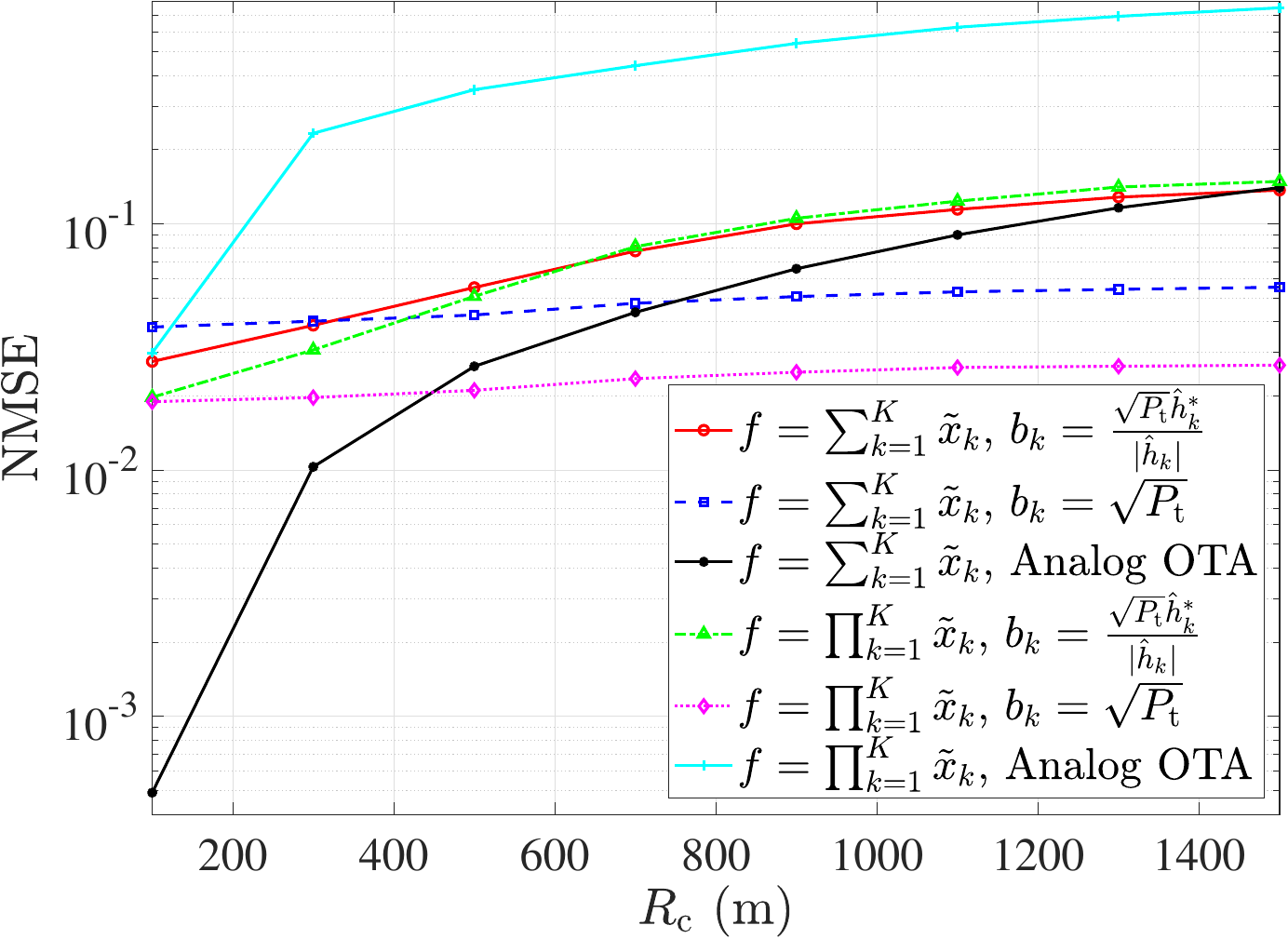}}
		}
		\subfigure[Symmetric function: maximum and sum of squares.]{
			\raisebox{0pt}[0pt][0pt]{\hspace*{0cm}}\includegraphics[width=0.45\linewidth]{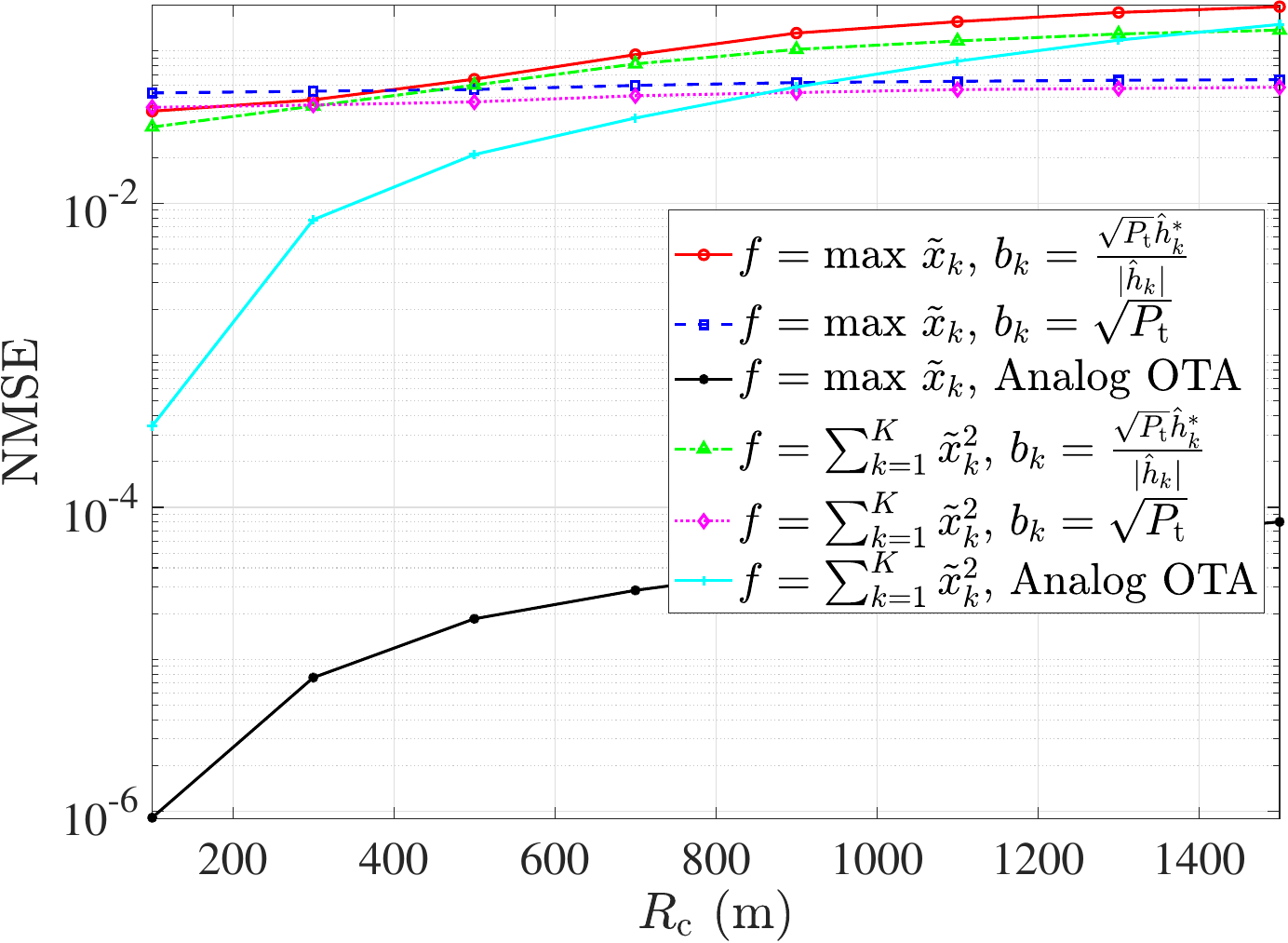}
		}
		
		\subfigure[Asymmetric function: sum and product.]{
			\raisebox{0pt}[0pt][0pt]{\hspace*{0cm}}{\includegraphics[width=0.45\linewidth]{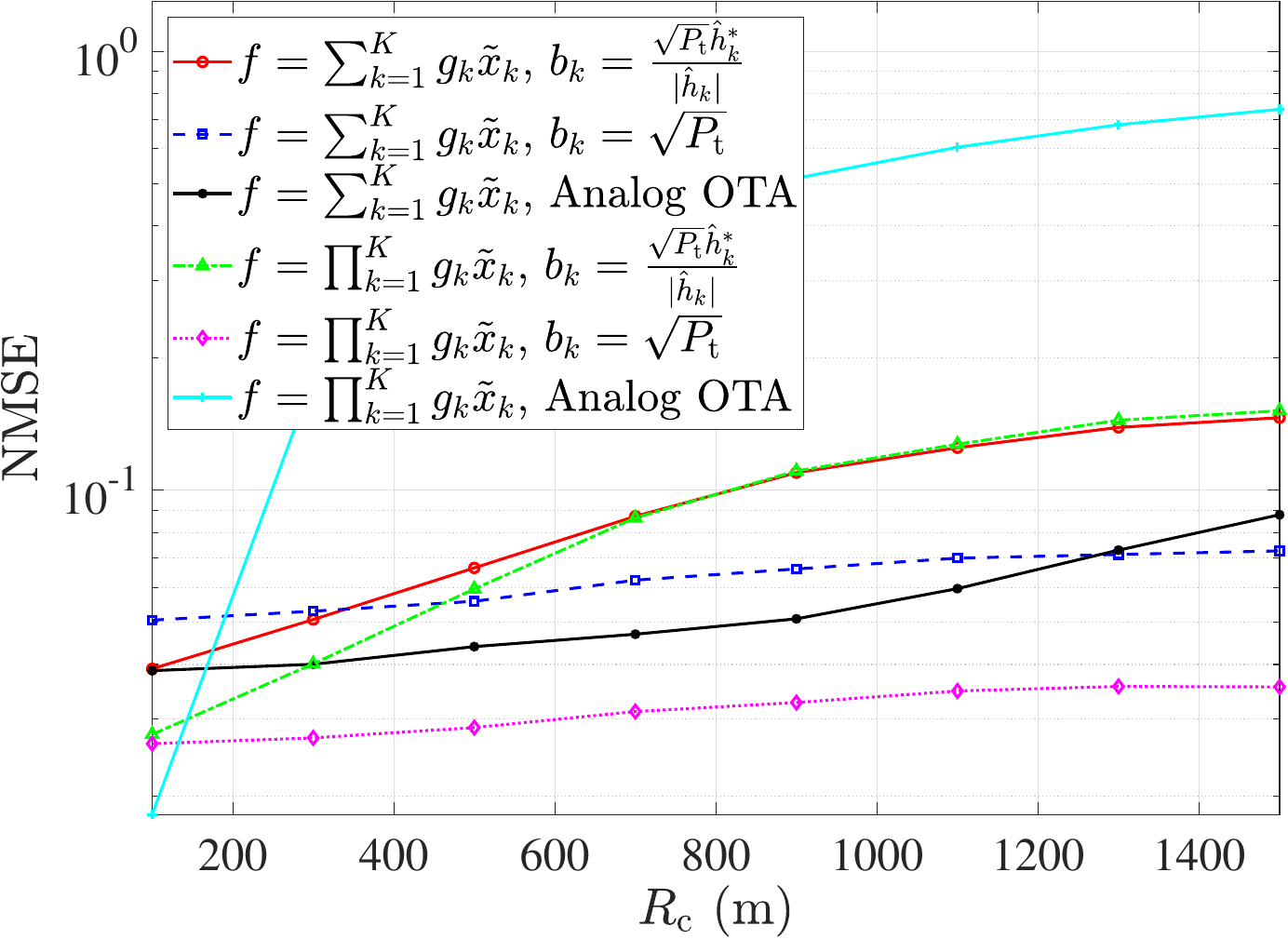}}
		}
		\subfigure[Asymmetric function: maximum and sum of squares.]{
			\raisebox{0pt}[0pt][0pt]{\hspace*{0cm}}{\includegraphics[width=0.45\linewidth]{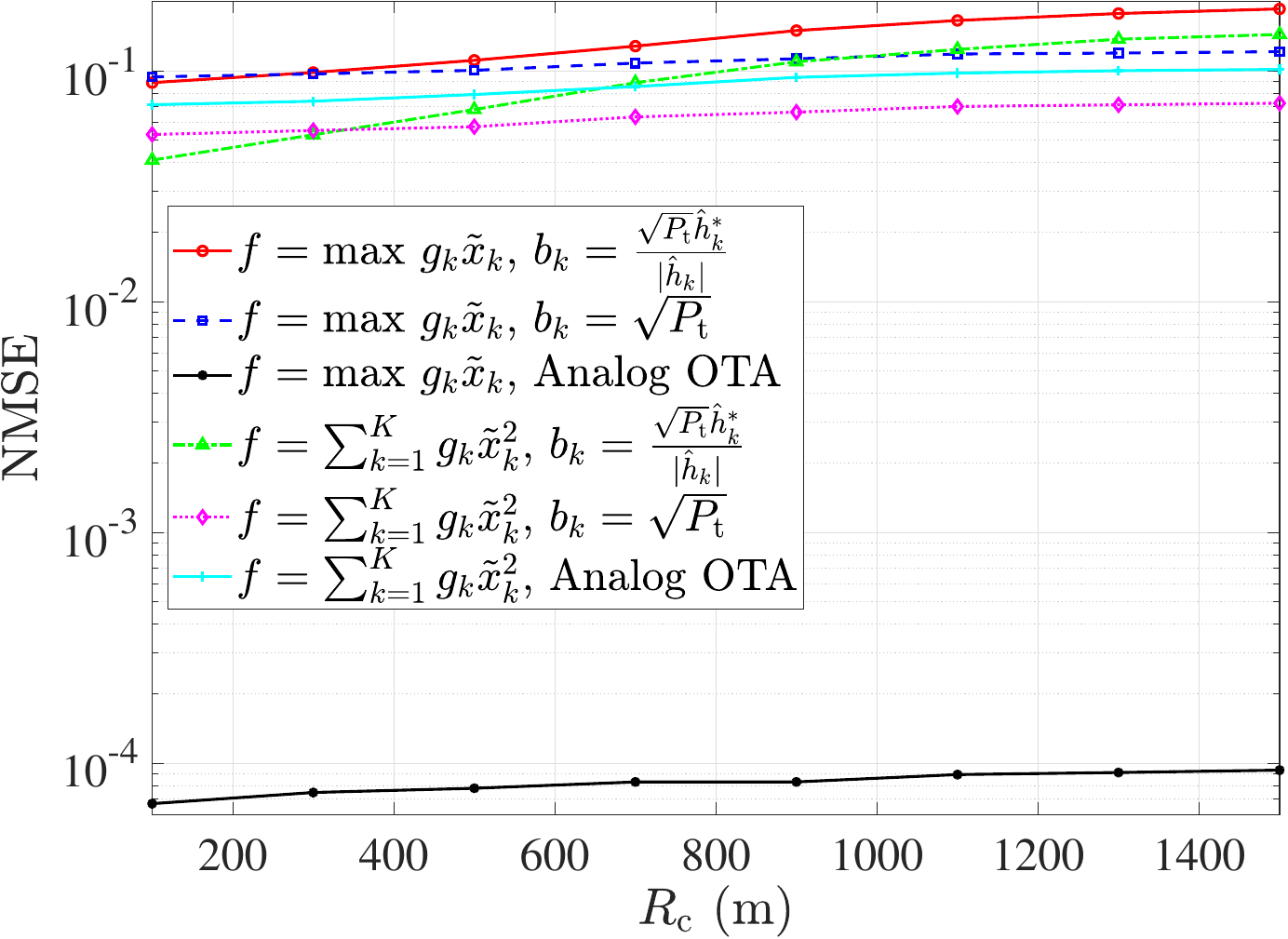}}
		}
		\caption{NMSE as a function of the radius of the cell $R_\mathrm{c}$ in cellular communication system. $K=4$, $Q=8$, $Q_1=3$, $Q_2=3$, $0 \leq \tilde{x}_k \leq 7$, $\breve{x}_k \in \{0, 1, 2, 3, 4, 5, 6, 7\}$, $g_k=k/K$.}
		\label{NMSE_D}  
	\end{figure*}
	
	First, the performance of channel-aware constellation for demodulation mapper in a cellular communication system is investigated. The CP with 144 antennas ($N_\mathrm{A}=144$) is located at the center of the cell while $K$ nodes (wireless devices) are uniformly distributed in the cell. The radius of the cell is denoted by $R_\mathrm{c}$. Fig.~\ref{NMSE_D} illustrates the NMSE performance as a function of the cell radius $R_c$ in a cellular communication system for both symmetric and asymmetric functions, highlighting the impact of different transmission coefficient ($b_k = \sqrt{P_t}$ and $b_k = \sqrt{P_\mathrm{t}} \hat{h}_k^* / |\hat{h}_k| $, where $\hat{h}_k=\sum_{n_\mathrm{A}=1}^{N_\mathrm{A}} \hat{h}_{k,n_\mathrm{A}}$). As shown in Fig.~\ref{NMSE_D}, the proposed channel aware constellation with $b_k = \sqrt{P_t}$ outperforms analog OTA in computing sum, product, and sum of squares functions as the cell radius $R_c$ increases, while analog OTA has lower NMSE in computing maximum. In addition, it indicates that $b_k = \sqrt{P_\mathrm{t}} \hat{h}_k^* / |\hat{h}_k|$ has lower NMSE than $b_k = \sqrt{P_t}$ only when $R_c$ is small. However, as $R_c$ increases, $b_k = \sqrt{P_t}$ outperforms $b_k = \sqrt{P_\mathrm{t}} \hat{h}_k^* / |\hat{h}_k|$. 
	The observed phenomenon can be attributed to the influence of \( R_\mathrm{c} \) on the signal-to-noise ratio (SNR) of individual nodes. When \( R_\mathrm{c} \) is small, the SNR at each node is relatively high. For the case where \( b_k = \sqrt{P_t} \), an example of the combined constellation utilized for demodulation is illustrated in Fig.~\ref{TransmitterandReceiverConstellation}(f). In this scenario, some symbol points representing distinct function values are located in close proximity to each other. This spatial closeness increases the susceptibility of these points to small noise perturbations. In contrast, for the case where \( b_k = \sqrt{P_\mathrm{t}} \hat{h}_k^* / |\hat{h}_k| \), the combined constellation for demodulation is depicted in Fig.~\ref{TransmitterandReceiverConstellation}(d). Here, the symbol points are distributed more uniformly across the constellation space. This uniform placement improves system performance under high SNR conditions, as no two points are positioned extremely close to each other. Therefore, \( b_k = \sqrt{P_t} \) results in a higher NMSE than \( b_k = \sqrt{P_\mathrm{t}} \hat{h}_k^* / |\hat{h}_k| \) under relatively high SNR. Conversely, the constellation of  \( b_k = \sqrt{P_t} \) also contains other symbol points that are separated by relatively larger distances, particularly those corresponding to different function values. These larger separations enhance robustness against low SNR conditions, reducing the likelihood of symbol misinterpretation. However, under low SNR conditions, for \( b_k = \sqrt{P_\mathrm{t}} \hat{h}_k^* / |\hat{h}_k| \), the uniform distribution of the constellation points---if limited in its overall spacing---may fail to maintain adequate separation between symbols. This reduction in separation degrades the reliability of demodulation, leading to a sharper increase in NMSE as SNR drops. Therefore, \( b_k = \sqrt{P_t} \) leads to a lower NMSE than \( b_k = \sqrt{P_\mathrm{t}} \hat{h}_k^* / |\hat{h}_k| \) under low SNR (when $R_\mathrm{c}$ is large).
	
	\begin{figure*}[t]
		\centering
		\subfigure[$f=\sum_{k=1}^K \breve{x}_k$.]{
			\raisebox{0pt}[0pt][0pt]{\hspace*{0cm}}{\includegraphics[width=0.23\linewidth]{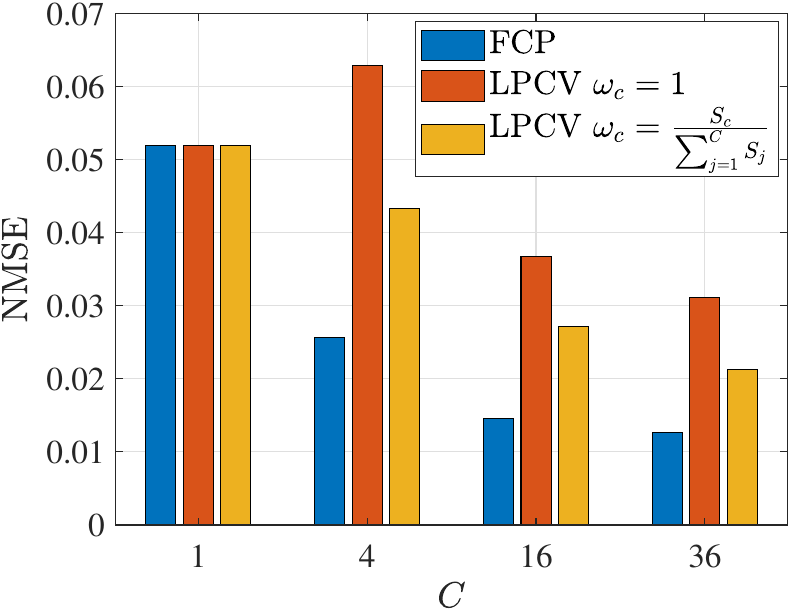}}
		}
		\subfigure[$f=\prod_{k=1}^K \breve{x}_k$.]{
			\raisebox{0pt}[0pt][0pt]{\hspace*{0cm}}\includegraphics[width=0.23\linewidth]{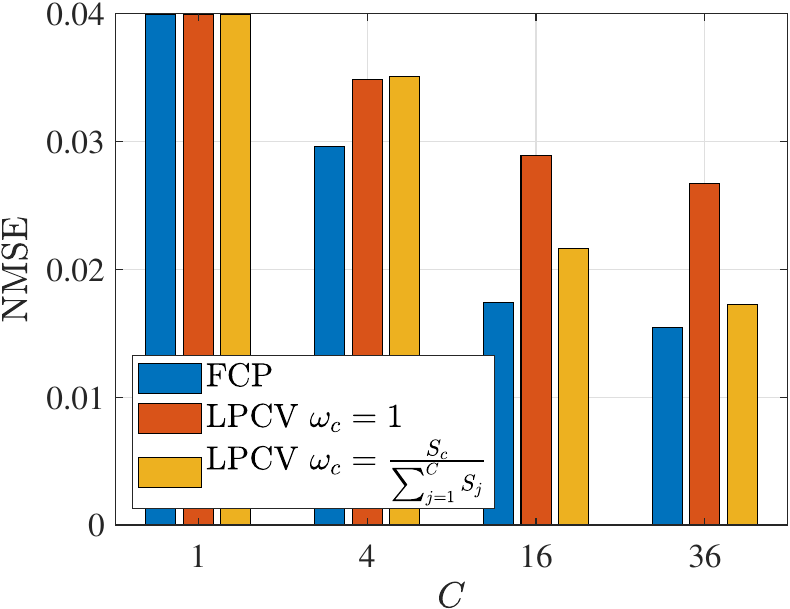}
		}
		\subfigure[$f=\text{max } \breve{x}_k$.]{
			\raisebox{0pt}[0pt][0pt]{\hspace*{0cm}}{\includegraphics[width=0.23\linewidth]{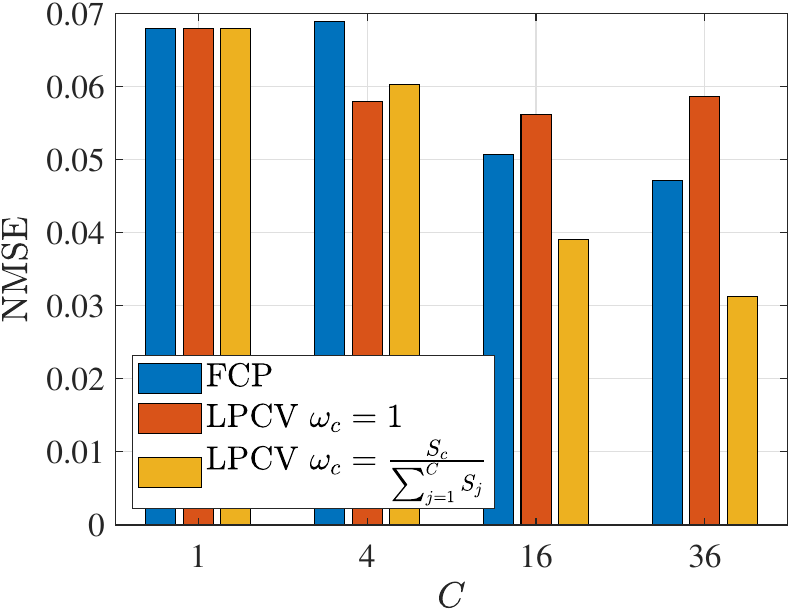}}
		}
		\subfigure[$f=\sum_{k=1}^K \breve{x}_k^2$.]{
			\raisebox{0pt}[0pt][0pt]{\hspace*{0cm}}{\includegraphics[width=0.23\linewidth]{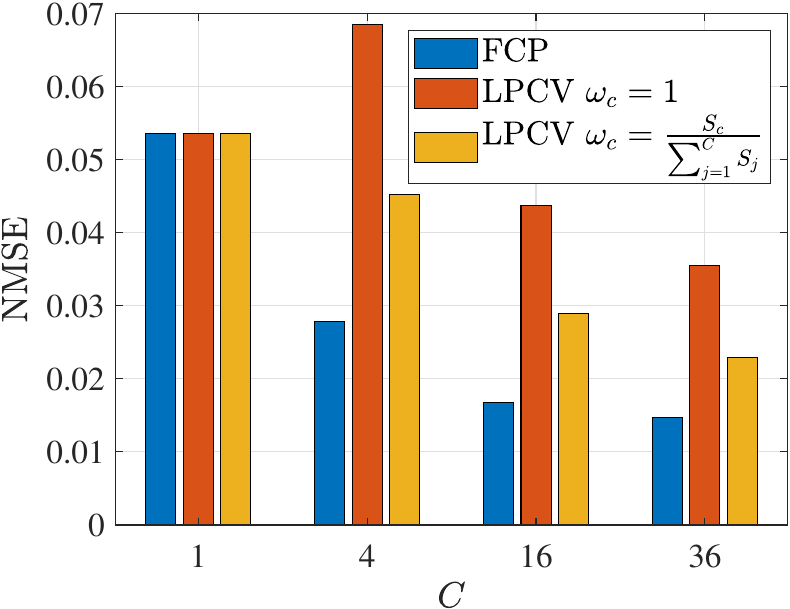}}
		}
		\caption{NMSE as a function of the number of CPs $C$ in cell-free communication system. $K=4$, $Q=4$,  $Q_1=2$, $Q_2=2$, $1 \leq \tilde{x}_k \leq 4$, $\breve{x}_k \in \{1, 2, 3, 4\}$, symmetric function.}
		\label{NMSEFunctions}  
	\end{figure*}
	
	\begin{figure*}[t]
		\centering
		\subfigure[$f=\sum_{k=1}^K g_k \breve{x}_k$.]{
			\raisebox{0pt}[0pt][0pt]{\hspace*{0cm}}{\includegraphics[width=0.23\linewidth]{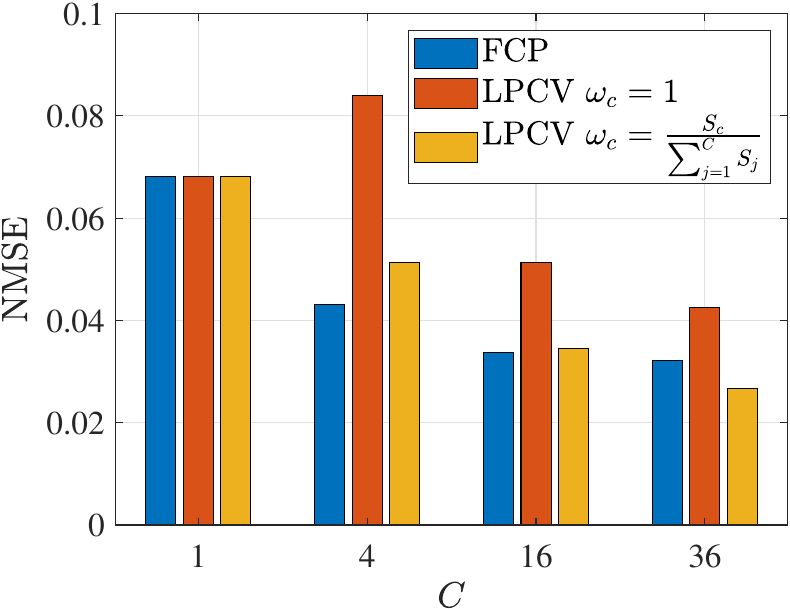}}
		}
		\subfigure[$f=\prod_{k=1}^K g_k \breve{x}_k$.]{
			\raisebox{0pt}[0pt][0pt]{\hspace*{0cm}}\includegraphics[width=0.23\linewidth]{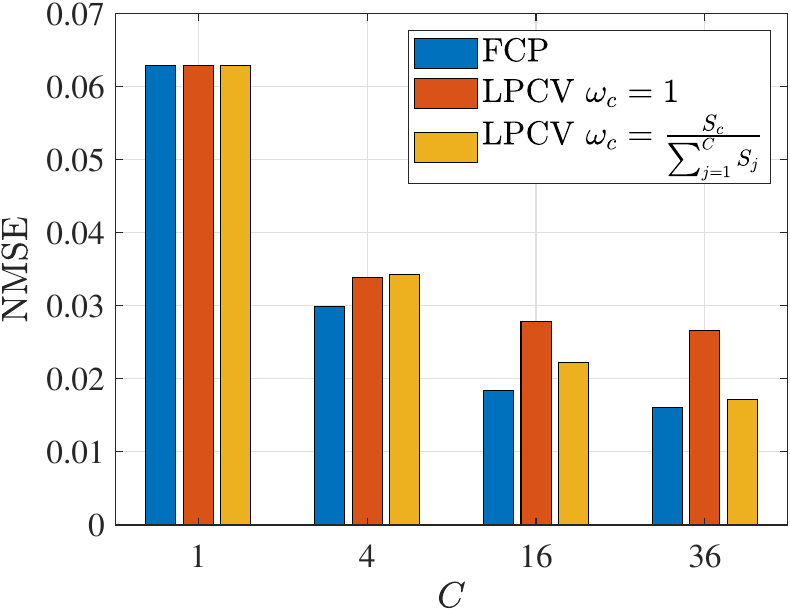}
		}
		\subfigure[$f=\text{max } g_k \breve{x}_k$.]{
			\raisebox{0pt}[0pt][0pt]{\hspace*{0cm}}{\includegraphics[width=0.23\linewidth]{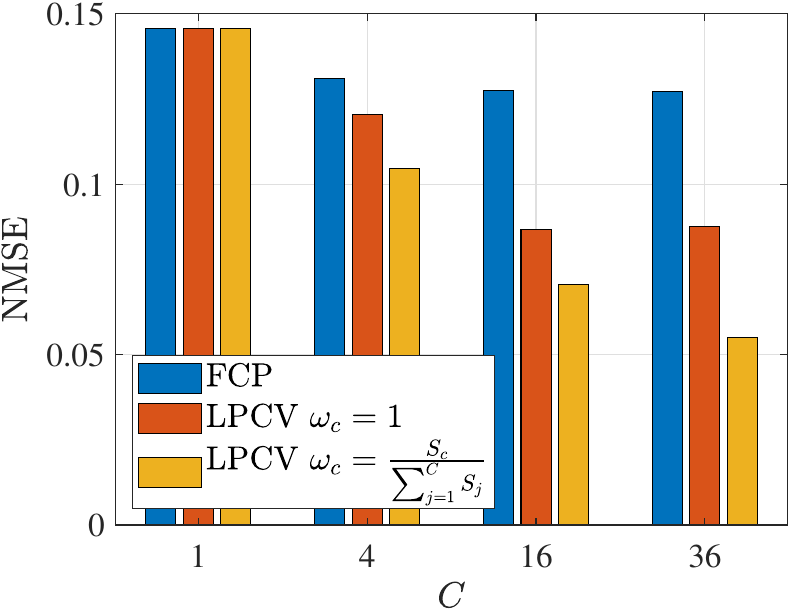}}
		}
		\subfigure[$f=\sum_{k=1}^K g_k \breve{x}_k^2$.]{
			\raisebox{0pt}[0pt][0pt]{\hspace*{0cm}}{\includegraphics[width=0.23\linewidth]{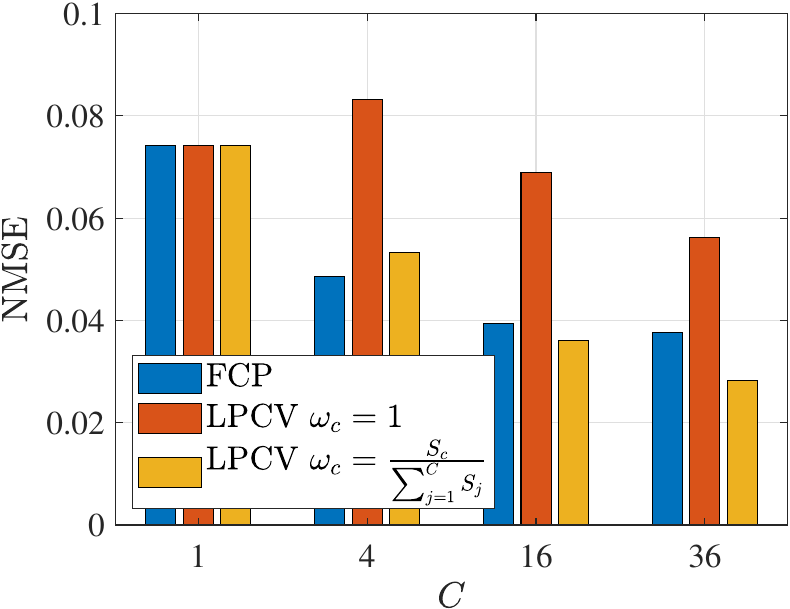}}
		}
		\caption{NMSE as a function of the number of CPs $C$ in cell-free communication system. $K=4$, $Q=4$, $Q_1=2$, $Q_2=2$,  $1 \leq \tilde{x}_k \leq 4$, $\breve{x}_k \in \{1, 2, 3, 4\}$, $g_k = k/K$, asymmetric function.}
		\label{NMSEAsymmFunctions}  
	\end{figure*}
	
	For cell-free communication system, we consider a $1 \times 1$ km simulation area. The CPs are deployed on a square grid. As comparison, a single base station equipped with 144 antennas is located in the center of the area. The total number of antennas in the Cell-free mMIMO network is set to 144 for fair comparison. $K$ nodes (wireless devices) are uniformly distributed in the considered area. Each CP is deployed with $144/K$ antennas. We adopt a random pilot assignment strategy and the uplink pilot powers are set to $P_\mathrm{t} = 20$ dBm. Fig.~\ref{NMSEFunctions} and Fig.~\ref{NMSEAsymmFunctions} present the performance of channel-aware constellation for demodulation mapper in a cell-free communication system. Specifically, Fig.~\ref{NMSEFunctions} presents symmetric functions (a) \( f = \sum_{k=1}^K \breve{x}_k \), (b) \( f = \prod_{k=1}^K \breve{x}_k \), (c) \( f = \max \breve{x}_k \), and (d) \( f = \sum_{k=1}^K \breve{x}_k^2 \), across varying number of CPs (\( C = 1, 4, 16, 36 \)), while Fig.~\ref{NMSEAsymmFunctions} presents asymmetric functions (a) \( f = \sum_{k=1}^K g_k \breve{x}_k \), (b) \( f = \prod_{k=1}^K g_k \breve{x}_k \), (c) \( f = \max g_k \breve{x}_k \), and (d) \( f = \sum_{k=1}^K g_k \breve{x}_k^2 \), across varying number of CPs (\( C = 1, 4, 16, 36 \)), where $g_k = k/K$. Three signal processing methods---FCP, LPCV with $\omega_c=1$, and LPCV with $\omega_c=\frac{S_i}{\sum_j S_j}$---are investigated. 
	As the number of CPs $C$ grows from 4 to 36, NMSE becomes lower. FCP generally achieves the best performance across most subfigures, as evidenced by consistently lower NMSE values. However, this comes at the cost of substantial fronthaul overhead from the CPs to the central CPU. In contrast, LPCV with $\omega_c = 1$ requires the least fronthaul overhead but tends to exhibit the highest NMSE among the three methods in most scenarios. The adaptive weighting approach, LPCV with $\omega_c = \tfrac{S_i}{\sum_j S_j}$, offers a compromise between these two extremes: it achieves lower NMSE than LPCV with $\omega_c = 1$ in most cases, and even the lowest NMSE for \( f = \max(\breve{x}_k) \) and \( f = \sum_{k=1}^K g_k \breve{x}_k^2 \) at \(C = 16\) and \(36\);\( f = \max(g_k\,\breve{x}_k) \) at \(C = 4, 16,\) and \(36\); and \( f = \sum_{k=1}^K g_k \,\breve{x}_k \) at \(C = 36\)---while also incurring lower fronthaul overhead than FCP.
	
	\begin{figure*}[t]
		\centering
		\subfigure[$f=\sum_{k=1}^K \breve{x}_k$.]{
			\raisebox{0pt}[0pt][0pt]{\hspace*{0cm}}{\includegraphics[width=0.23\linewidth]{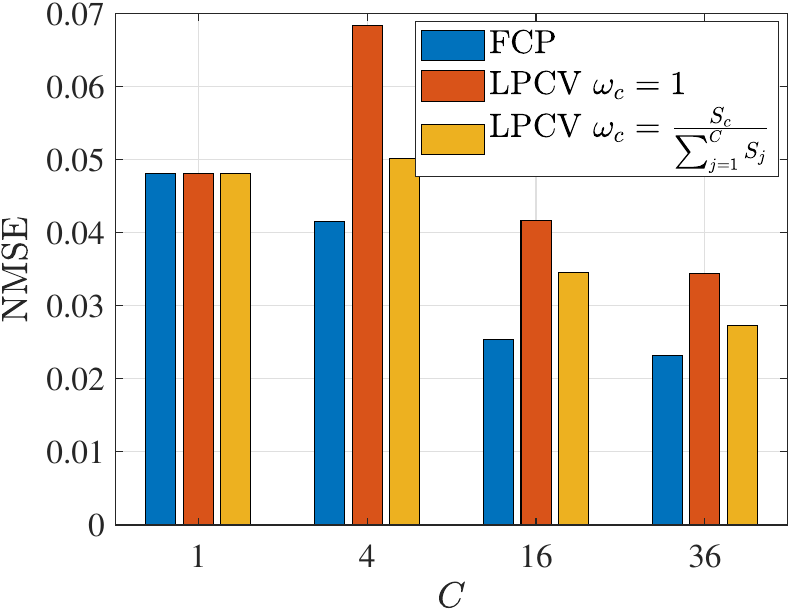}}
		}
		\subfigure[$f=\prod_{k=1}^K \breve{x}_k$.]{
			\raisebox{0pt}[0pt][0pt]{\hspace*{0cm}}\includegraphics[width=0.23\linewidth]{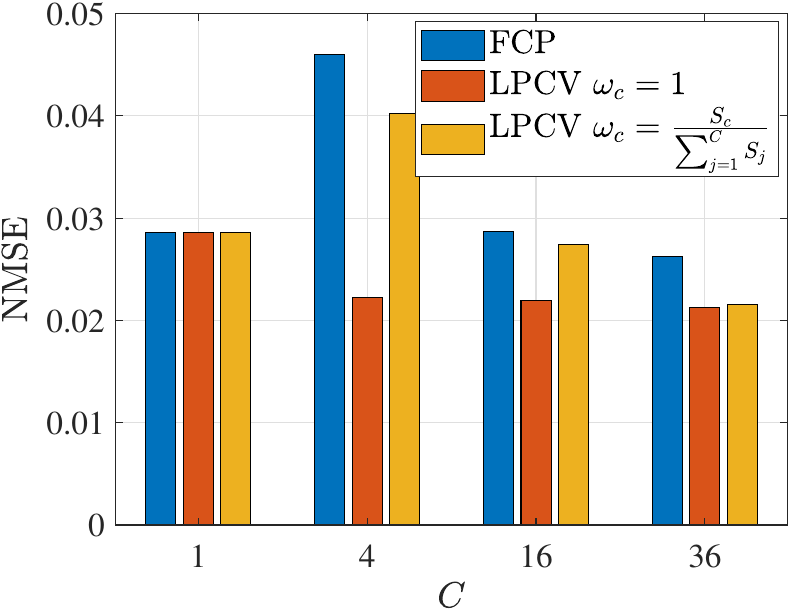}
		}
		\subfigure[$f=\text{max } \breve{x}_k$.]{
			\raisebox{0pt}[0pt][0pt]{\hspace*{0cm}}{\includegraphics[width=0.23\linewidth]{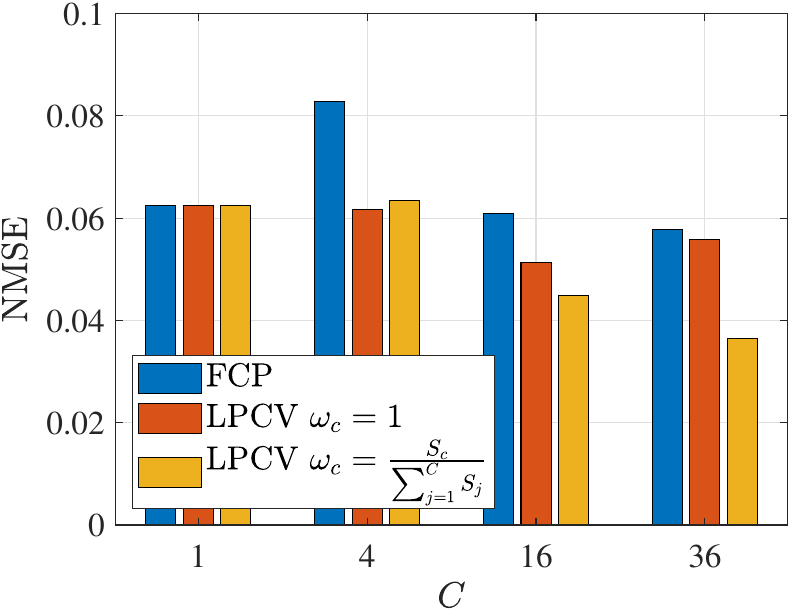}}
		}
		\subfigure[$f=\sum_{k=1}^K \breve{x}_k^2$.]{
			\raisebox{0pt}[0pt][0pt]{\hspace*{0cm}}{\includegraphics[width=0.23\linewidth]{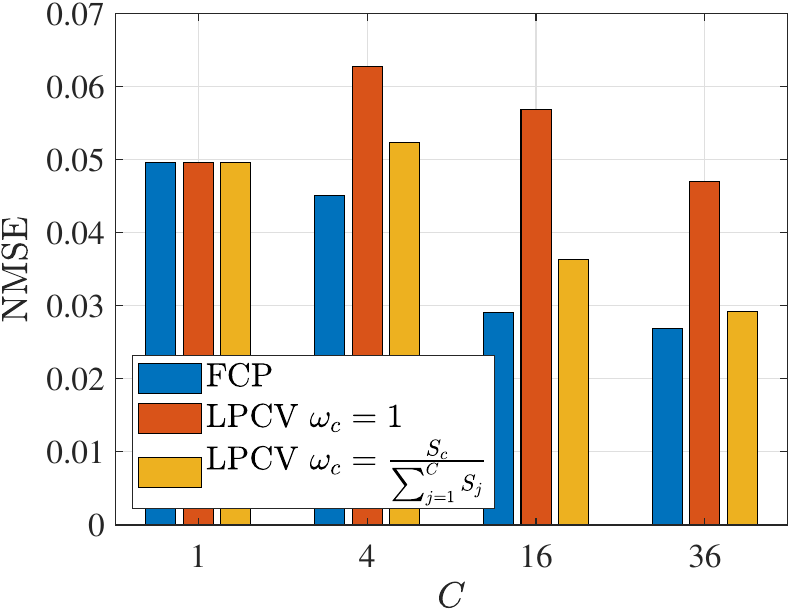}}
		}
		\caption{NMSE as a function of the number of CPs $C$ in cell-free communication system. $K=4$, $Q=8$, $Q_1=3$, $Q_2=3$, $1 \leq \tilde{x}_k \leq 8$, $\breve{x}_k \in \{1, 2, 3, 4, 5, 6, 7, 8\}$, symmetric function.}
		\label{NMSEFunctions2}    
	\end{figure*}
	
	\begin{figure*}[t]
		\centering
		\subfigure[$f=\sum_{k=1}^K g_k \breve{x}_k$.]{
			\raisebox{0pt}[0pt][0pt]{\hspace*{0cm}}{\includegraphics[width=0.23\linewidth]{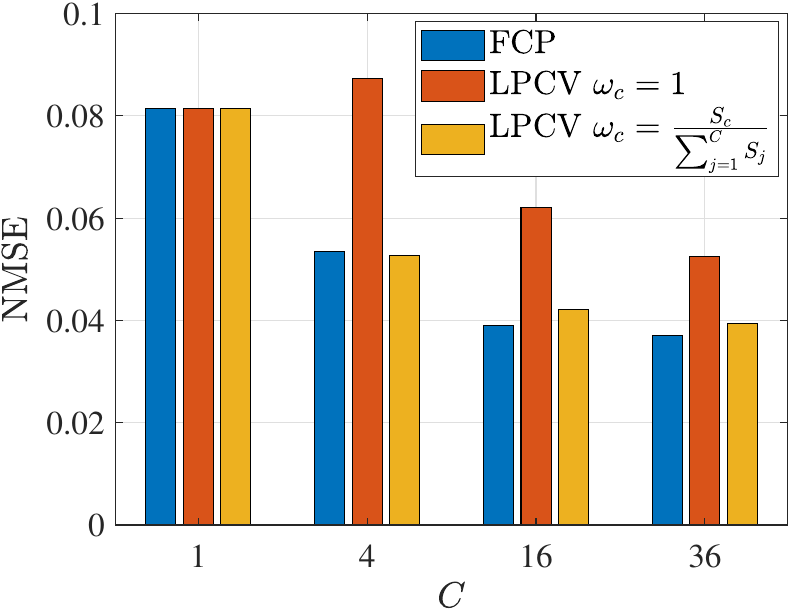}}
		}
		\subfigure[$f=\prod_{k=1}^K g_k \breve{x}_k$.]{
			\raisebox{0pt}[0pt][0pt]{\hspace*{0cm}}\includegraphics[width=0.23\linewidth]{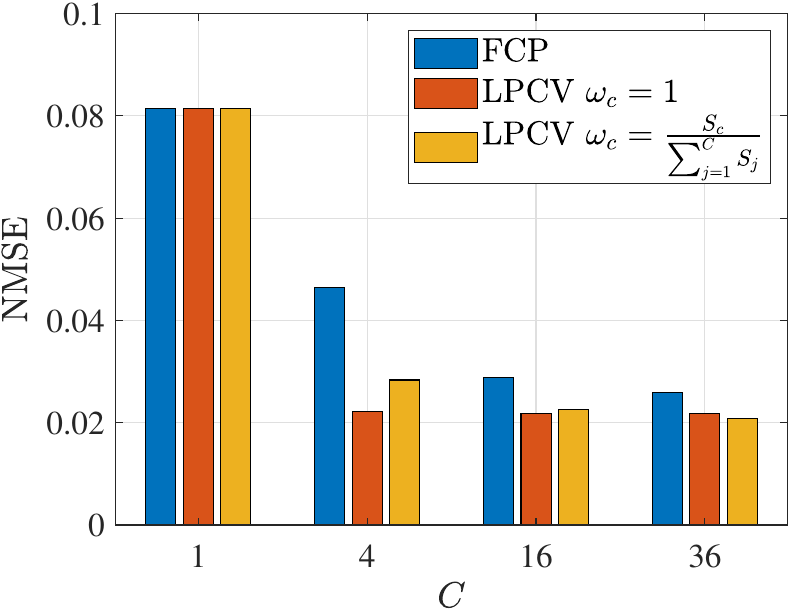}
		}
		\subfigure[$f=\text{max } g_k \breve{x}_k$.]{
			\raisebox{0pt}[0pt][0pt]{\hspace*{0cm}}{\includegraphics[width=0.23\linewidth]{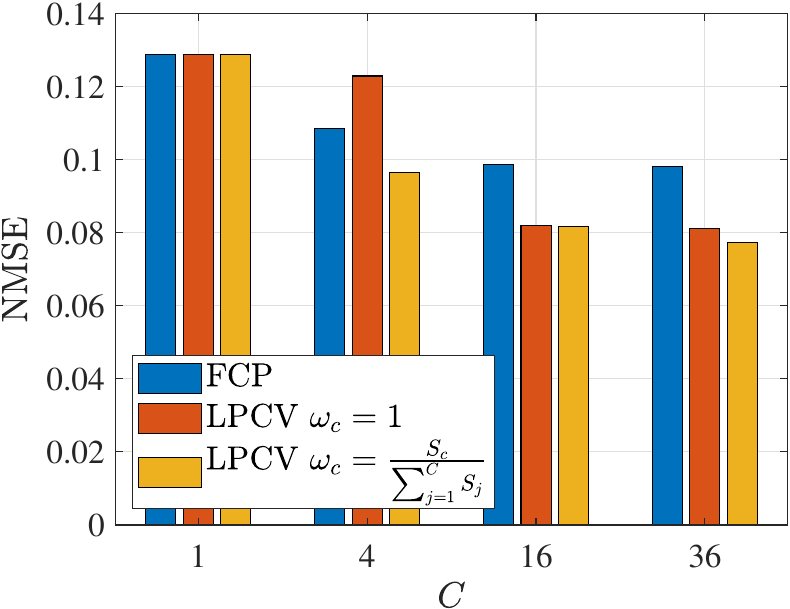}}
		}
		\subfigure[$f=\sum_{k=1}^K g_k \breve{x}_k^2$.]{
			\raisebox{0pt}[0pt][0pt]{\hspace*{0cm}}{\includegraphics[width=0.23\linewidth]{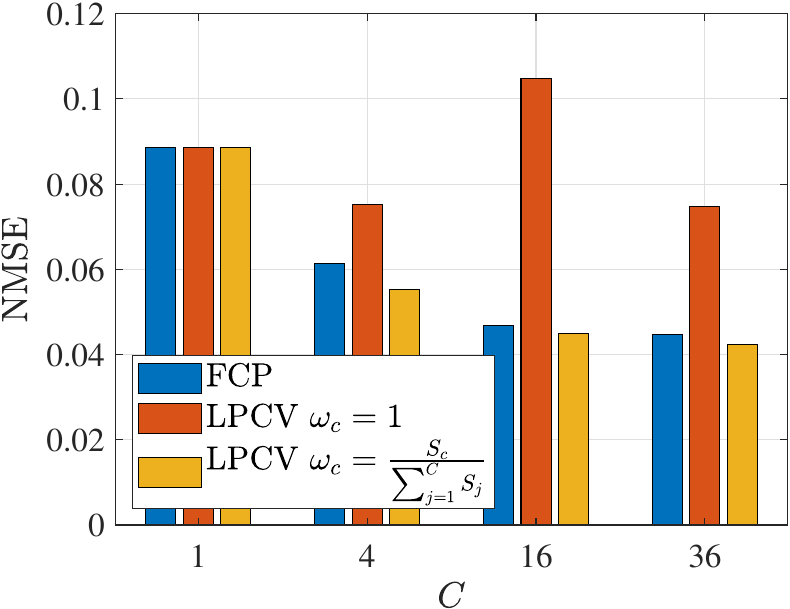}}
		}
		\caption{NMSE as a function of the number of CPs $C$ in cell-free communication system. $K=4$, $Q=8$, $Q_1=3$, $Q_2=3$, $1 \leq \tilde{x}_k \leq 8$, $\breve{x}_k \in \{1, 2, 3, 4, 5, 6, 7, 8\}$, $g_k = k/K$, asymmetric function.}
		\label{NMSEAsymmFunctions2}  
	\end{figure*}
	
	Fig.~\ref{NMSEFunctions2} and Fig.~\ref{NMSEAsymmFunctions2} extend the analysis presented in Fig.~\ref{NMSEFunctions} and Fig.~\ref{NMSEAsymmFunctions} by increasing the number of possible constellation symbols from \( Q = 4 \) to \( Q = 8 \), resulting in \( \breve{x}_k \in \{1, 2, 3, 4, 5, 6, 7, 8\} \). This increase in constellation size introduces higher complexity, which generally leads to slightly higher NMSE values across the most functions compared to Fig.~\ref{NMSEFunctions} and Fig.~\ref{NMSEAsymmFunctions}. Despite this, the trends remain consistent: FCP achieves the lowest NMSE but incurs the highest fronthaul overhead, LPCV with \( \omega_c = 1 \) exhibits the highest NMSE with minimal fronthaul overhead, and LPCV with \( \omega_c = \frac{S_i}{\sum_j S_j} \) provides a balanced compromise. The impact of increased \( Q \) is particularly evident for product-based functions such as \( f = \prod_{k=1}^K \breve{x}_k \) and \( f = \prod_{k=1}^K g_k \breve{x}_k \), where NMSE differences among methods become more pronounced at larger \( C \). Additionally, LPCV with \( \omega_c = \frac{S_i}{\sum_j S_j} \) demonstrates improved robustness under increased constellation complexity, achieving the lowest NMSE in cases such as \( f = \max \breve{x}_k \) and \( f = \sum_{k=1}^K g_k \breve{x}_k^2 \) for higher \( C \) values. Overall, while Fig.~\ref{NMSEFunctions2} and Fig.~\ref{NMSEAsymmFunctions2} exhibit slightly higher NMSE values due to the increased constellation size, the relative performance of the methods and their trade-offs remain consistent with the observations from Fig.~\ref{NMSEFunctions} and Fig.~\ref{NMSEAsymmFunctions}.
	
	\begin{figure*}[t]
		\centering
		\subfigure[$f=\sum_{k=1}^K \breve{x}_k$.]{
			\raisebox{0pt}[0pt][0pt]{\hspace*{0cm}}{\includegraphics[width=0.23\linewidth]{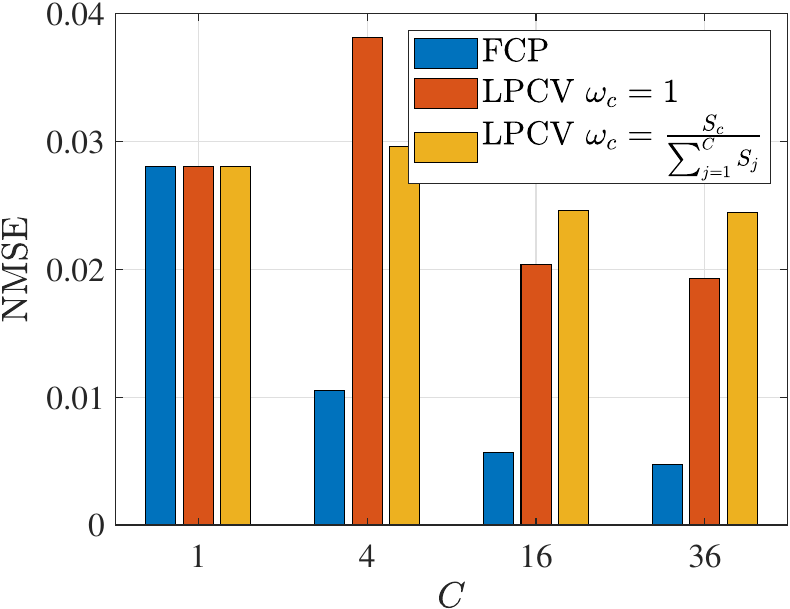}}
		}
		\subfigure[$f=\prod_{k=1}^K \breve{x}_k$.]{
			\raisebox{0pt}[0pt][0pt]{\hspace*{0cm}}\includegraphics[width=0.23\linewidth]{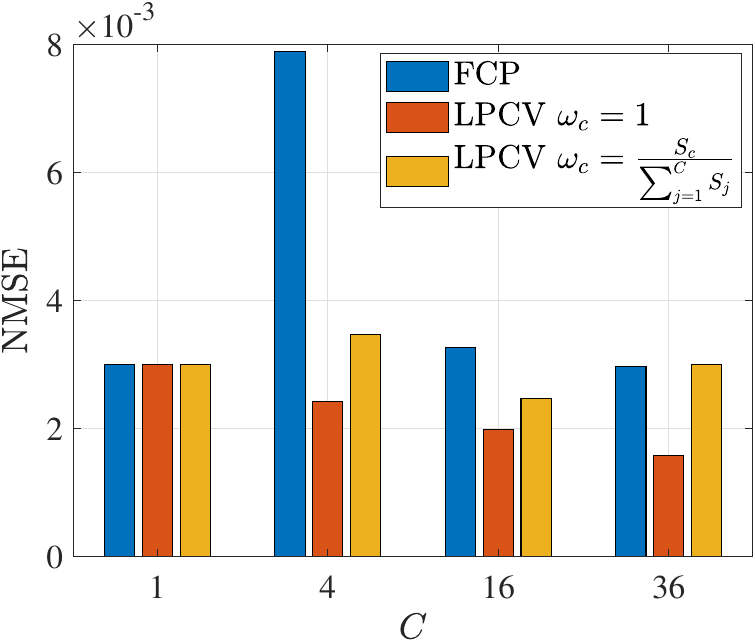}
		}
		\subfigure[$f=\text{max } \breve{x}_k$.]{
			\raisebox{0pt}[0pt][0pt]{\hspace*{0cm}}{\includegraphics[width=0.23\linewidth]{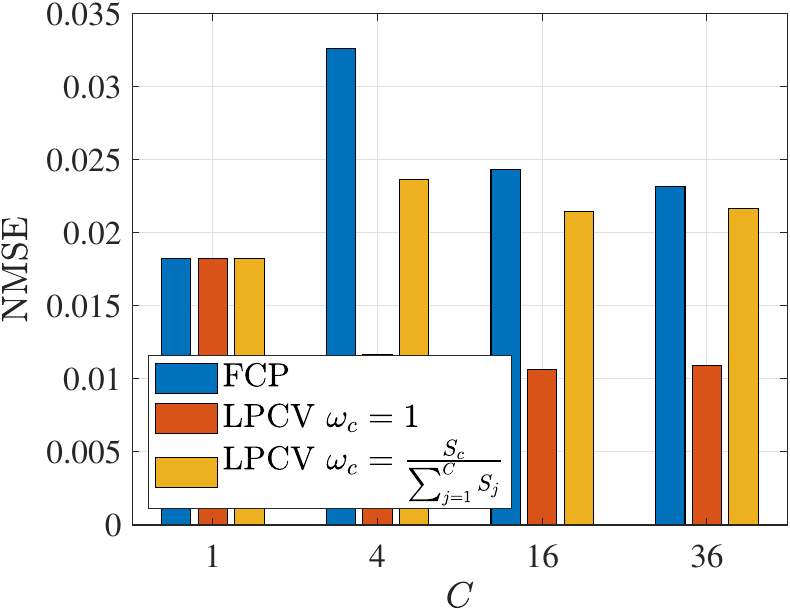}}
		}
		\subfigure[$f=\sum_{k=1}^K \breve{x}_k^2$.]{
			\raisebox{0pt}[0pt][0pt]{\hspace*{0cm}}{\includegraphics[width=0.23\linewidth]{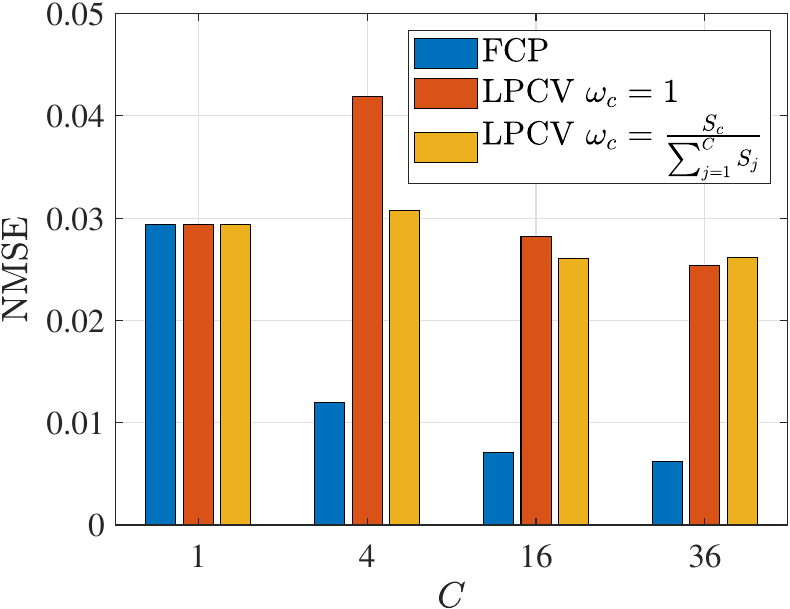}}
		}
		\caption{NMSE as a function of the number of CPs $C$ in cell-free communication system. $K=8$, $Q=4$, $Q_1=2$, $Q_2=2$, $1 \leq \tilde{x}_k \leq 4$, $\breve{x}_k \in \{1, 2, 3, 4\}$, symmetric function.}
		\label{NMSEFunctions3}    
	\end{figure*}
	
	\begin{figure*}[t]
		\centering
		\subfigure[$f=\sum_{k=1}^K g_k \breve{x}_k$.]{
			\raisebox{0pt}[0pt][0pt]{\hspace*{0cm}}{\includegraphics[width=0.23\linewidth]{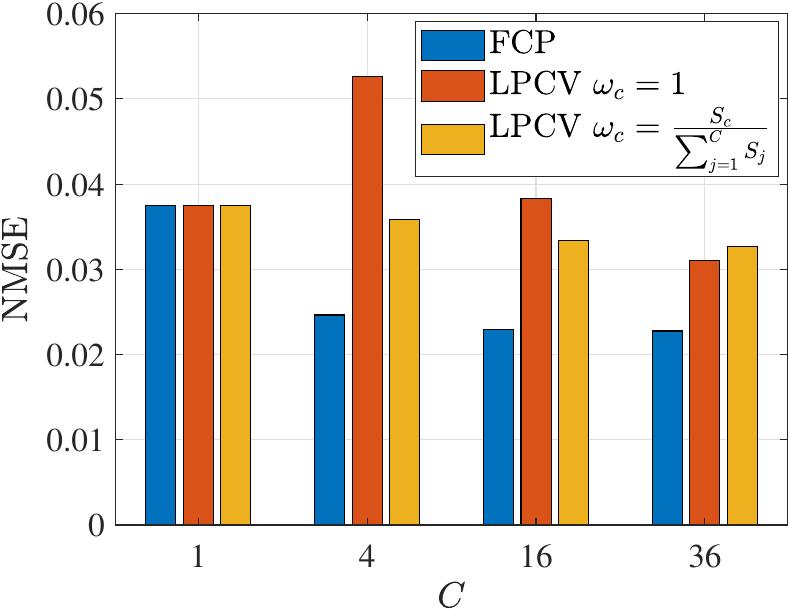}}
		}
		\subfigure[$f=\prod_{k=1}^K g_k \breve{x}_k$.]{
			\raisebox{0pt}[0pt][0pt]{\hspace*{0cm}}\includegraphics[width=0.23\linewidth]{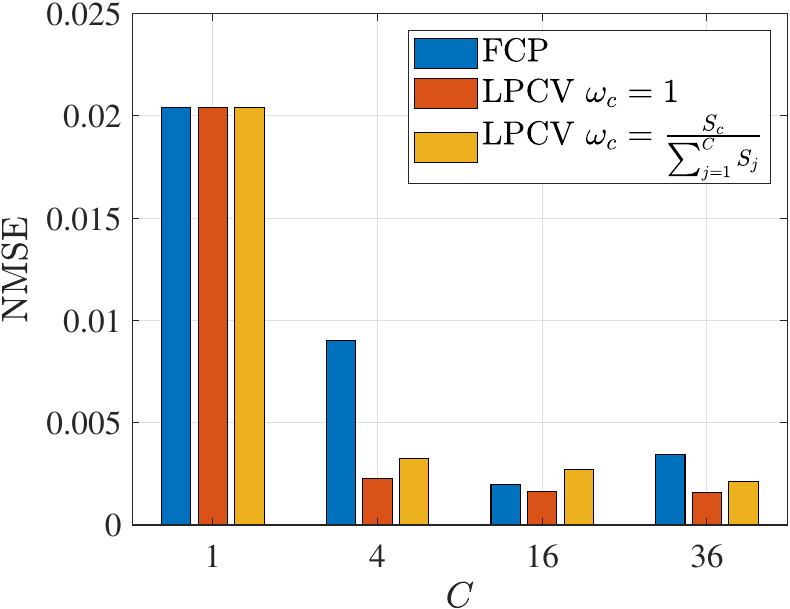}
		}
		\subfigure[$f=\text{max } g_k \breve{x}_k$.]{
			\raisebox{0pt}[0pt][0pt]{\hspace*{0cm}}{\includegraphics[width=0.23\linewidth]{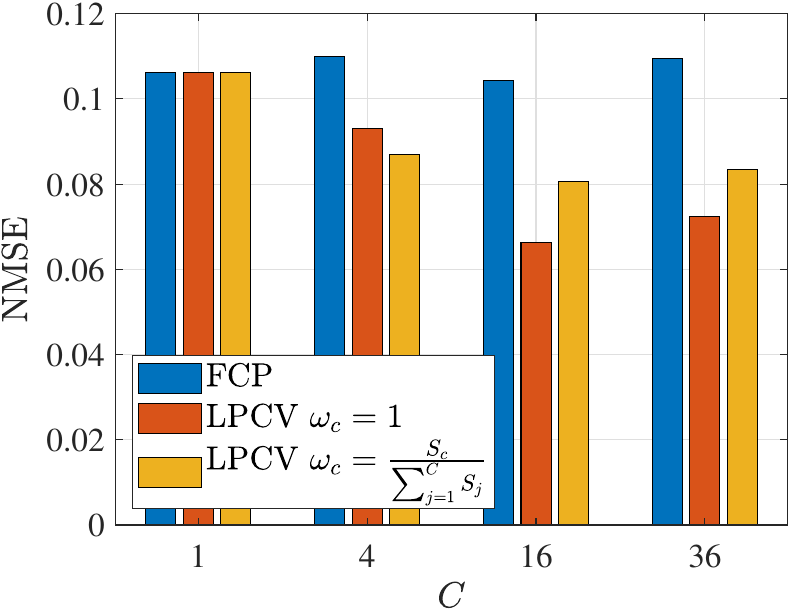}}
		}
		\subfigure[$f=\sum_{k=1}^K g_k \breve{x}_k^2$.]{
			\raisebox{0pt}[0pt][0pt]{\hspace*{0cm}}{\includegraphics[width=0.23\linewidth]{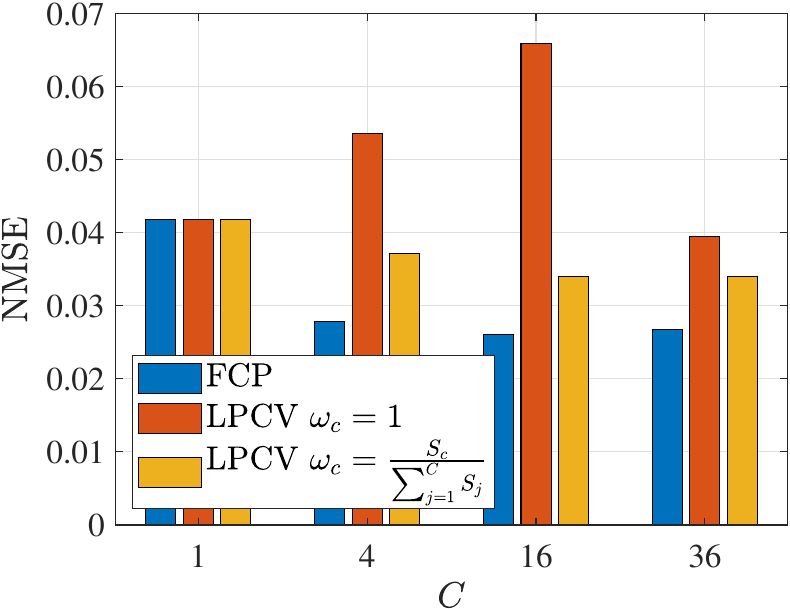}}
		}
		\caption{NMSE as a function of the number of CPs $C$ in cell-free communication system. $K=8$, $Q=4$, $Q_1=2$, $Q_2=2$, $1 \leq \tilde{x}_k \leq 4$, $\breve{x}_k \in \{1, 2, 3, 4\}$, $g_k = k/K$, asymmetric function.}
		\label{NMSEAsymmFunctions3}  
	\end{figure*}
	
	Fig.~\ref{NMSEFunctions3} and Fig.~\ref{NMSEAsymmFunctions3} provide a new perspective by considering \( K = 8 \) and \( Q = 4 \). From the figures, in many cases, FCP still demonstrates competitive or even superior NMSE performance. Interestingly, when \( K = 8 \), LPCV with \( \omega_c = 1 \) outperforms LPCV with \( \omega_c = \frac{S_i}{\sum_j S_j} \) in certain scenarios. Specifically, in Fig.~\ref{NMSEFunctions3}, for symmetric functions such as \( f = \prod_{k=1}^K \breve{x}_k \) and \( f = \max g_k \breve{x}_k \), LPCV with \( \omega_c = 1 \) achieves lowest NMSE at \( C = 4 \), \( 16 \) and \( 36 \). Similarly, in Fig.~\ref{NMSEAsymmFunctions3}, for asymmetric functions such as \( f = \prod_{k=1}^K g_k \breve{x}_k \) and \( f = \max g_k \breve{x}_k \), LPCV with \( \omega_c = 1 \) achieves lowest NMSE performance at \( C = 4 \), \( 16 \) and \( 36 \). For instance, as the number of CPs \( C \) increases, LPCV’s simpler uniform weighting strategy (\(\omega_c = 1\)) can sometimes yield equally low or lower NMSE. 
	This suggests that, although FCP often excels, the performance gap between FCP and LPCV may diminish under certain objective functions or larger-scale settings. Overall, these observations confirm that neither approach strictly dominates across all conditions and underscore the importance of selecting the appropriate method and weighting strategy based on the specific requirements of the system and the nature of the function. 
	
	\section{Conclusion}
	In this paper, we proposed a digital OTA computation system with a channel-aware constellation design for demodulation mappers, applicable to both cellular and cell-free massive MIMO systems. The proposed system dynamically adapts the constellation based on channel conditions, reducing computational complexity while ensuring reliable function estimation. By accommodating both symmetric and asymmetric functions, the system significantly broadens the range of potential applications beyond traditional OTA approaches. The numerical results demonstrated the effectiveness of the proposed system in achieving low normalized mean square error. While fully centralized processing consistently achieves the lowest normalized mean square error, its high fronthaul overhead limits its scalability. In contrast, local processing $\&$ centralized voting approach with adaptive weighting offers a balanced trade-off between performance and overhead, making it a viable alternative for practical deployment in large-scale systems. Overall, the proposed channel-aware constellation system addresses key challenges in digital OTA computation, including excessive power consumption of wireless devices, computational complexity, and adaptability to diverse communication environments. Future research could explore the integration of advanced machine learning techniques for further optimization.
	
	\appendices
	\section{Proof of Proposition~\ref{proposition1}}
	Let \(\Tilde{s}_{m_1}\) and \(\Tilde{s}_{m_2}\) be the constellation points corresponding to 
	two distinct symbols (or messages) \(m_1\) and \(m_2\). Specifically,
	\begin{equation*}
		\begin{aligned}
			\Tilde{s}_{m_1} &
			=
			\sqrt{P_\mathrm{t}} \text{Sum} [|h_1| x_{1, q_{1,m_1}},\, |h_2| x_{2, q_{2,m_1}},\dots, |h_k| x_{K,q_{K,m_1}} ], \\
			\Tilde{s}_{m_2} &
			= \sqrt{P_\mathrm{t}} \text{Sum} [|h_1| x_{1, q_{1,m_2}},\, |h_2| x_{2, q_{2,m_2}},\dots, |h_k| x_{K,q_{K,m_2}} ].
		\end{aligned}
	\end{equation*}
	Since \(m_1 \neq m_2\), there is at least one index \(k\) such that
	\(x_{k,q_{k,m_1}} \neq x_{k,q_{k,m_2}}\).
	
	Assume the channel gains \(\{h_k\}\) are i.i.d.\ circularly symmetric complex 
	Gaussian with zero mean and variance \(\sigma^2\). Consequently, each 
	\(\lvert h_k \rvert\) follows a Rayleigh distribution with a continuous 
	probability density function.  
	
	We focus on the event in which two different transmit vectors, 
	\(x_{m_1}\) and \(x_{m_2}\), yield the same \emph{composite} received signal 
	(neglecting additive noise for simplicity).  In mathematical form, this overlap 
	requires
	\begin{equation}
		\sum_{k=1}^K \lvert h_k \rvert \, x_{k,q_{k,m_1}}
		\;=\;
		\sum_{k=1}^K \lvert h_k \rvert \, x_{k,q_{k,m_2}}.
		\label{Proof1}
	\end{equation}
	Let
	\[
	\Delta_k 
	\;=\; 
	x_{k,q_{k,m_1}} \;-\; x_{k,q_{k,m_2}}.
	\]
	Because \(x_{m_1}\) and \(x_{m_2}\) are different, there exists at least one 
	\(k\) for which \(\Delta_k \neq 0\).  Then Eq. \eqref{Proof1} is equivalent to
	\begin{equation}
		\sum_{k=1}^K \Delta_k \,\lvert h_k \rvert
		\;=\;
		0.
		\label{Proof2}
	\end{equation}
	
	Since each \(\lvert h_k \rvert\) is a continuous random variable (under the Rayleigh distribution) and they are independent across \(k\), the collection of points \((\lvert h_1 \rvert, \dots, \lvert h_K \rvert)\) satisfying Eq. \eqref{Proof2} lies on a \((K-1)\)-dimensional hyperplane in \(\mathbb{R}^K\). From measure theory, the probability of randomly drawing a point in a continuous space that falls exactly on a lower-dimensional hyperplane is zero \cite{B1}.  
	
	An analogous argument holds if one uses the full complex coefficients \(h_k\) (not just their magnitudes).  In that case, the condition becomes \(\sum_{k=1}^K \Delta_k\,h_k = 0\), which again describes a lower-dimensional subset in the continuous space of channel realizations \(\mathbb{C}^K\). Hence, the probability is still zero.
	
	Therefore, for two distinct symbols \(m_1 \neq m_2\), the probability of their corresponding constellation points $\Tilde{s}_{m_1}$ (or $\Breve{s}_{m_1}$) and $\Tilde{s}_{m_2}$ (or $\Breve{s}_{m_2}$) overlapping after Rayleigh fading (i.e., satisfying the above sum-equality) is zero.  In other words, under random fading with continuous distributions, different transmit signal vectors almost surely remain distinguishable at the receiver.
	
	\section{Proof of Proposition~\ref{prop:unique_combined_constellation}}
	First, note that each symbol $m$ corresponds to a distinct transmit vector \(\{ x_{k,q_{k,m}} \}_{k=1}^K\).  For Rayleigh or other continuous fading distributions, the channel gains $\{h_k\}$ (or their magnitudes $\{|h_k|\}$) are almost surely distinct in a measure-theoretic sense. Hence, the linear combinations of the form in~\eqref{eq:constellation_combination} map each $\{x_{k,q_{k,m}}\}$ to a unique point in the complex (or real) signal space, denoted by $\tilde{s}_m$ or $\breve{s}_m$.
	
	To see why the mapping is injective (one-to-one), suppose two different symbol indices $m_1 \neq m_2$ yielded the same combined constellation point, i.e.,
	\begin{align*}
		\sum_{k=1}^{K} \sqrt{P_\mathrm{t}}\,|h_k|\,x_{k,q_{k,m_1}}
		\;=\;
		\sum_{k=1}^{K} \sqrt{P_\mathrm{t}}\,|h_k|\,x_{k,q_{k,m_2}},
	\end{align*}
	(and similarly for $h_k$ instead of $|h_k|$). By rearranging terms, one would find that this requires a linear combination of independent continuous random variables (the channel gains) to be identically zero in a way that contradicts the distinctness of $x_{k,q_{k,m_1}}$ and $x_{k,q_{k,m_2}}$. Therefore, the probability of such an exact overlap is zero, proving injectivity almost surely.
	
	Because the mapping is injective, different symbols map to distinct points $\tilde{s}_m$ or $\breve{s}_m$, thus ensuring that symmetric or asymmetric function definitions (e.g., different coding/decoding strategies) can be supported by appropriately choosing the transmit vectors $\{x_{k,q_{k,m}}\}$. This flexibility stems from the fact that each distinct set of transmit waveforms experiences a unique fade combination, giving rise to a unique constellation point at the CP.
	
	Consequently, \eqref{eq:constellation_combination} defines a one-to-one mapping into the combined constellation space, allowing both symmetric and asymmetric functional designs to coexist in the system.


\begin{thebibliography}{1}
		
		\bibitem{A1} H. Hellstrom, J. M. B. da Silva Jr, M. M. Amiri, M. Chen, V. Fodor, H. V. Poor, C. Fischione, ``Wireless for machine learning: A survey," \textit{Foundations and Trends in Signal Processing}, vol. 15, no. 4,
		pp. 290–399, 2022.
		
		\bibitem{A2} H. Tataria, M. Shaf, A. F. Molisch, M. Dohler, H. Sjoland, and F. Tufvesson, ``6G wireless systems: Vision, requirements, challenges, insights, and opportunities," \textit{Proc. IEEE}, vol. 109, no. 7, pp. 1166–1199, 2021.
		
		\bibitem{A3} Z. Chen, E. G. Larsson, C. Fischione, M. Johansson, and Y. Malitsky, ``Over-the-air computation for distributed systems: Something old and something new," \textit{IEEE Netw.}, vol. 37, no. 5, pp. 240-246, Sept. 2023.
		
		\bibitem{A4} A. S¸ahin and R. Yang, ``A survey on over-the-air computation," \textit{IEEE
			Commun. Surv. Tutor.}, vol. 25, no. 3, pp. 1877-1908, thirdquarter 2023.
		
		\bibitem{A5} B. Nazer and M. Gastpar, ``Computation over multiple-access channels," \textit{IEEE Trans. Info. Theo.}, vol. 53, no. 10, pp. 3498–3516, Oct. 2007.
		
		\bibitem{A5_2} G. Zhu, Y. Du, D. Gunduz, and K. Huang, ``One-bit over-the-air aggregation for communication-efficient federated edge learning: Design and convergence analysis," \textit{IEEE Trans. Wireless Commun.}, vol. 20,no. 3, pp. 2120–2135, Mar. 2021.
		
		\bibitem{A5_3} A. Şahin, B. Everette, and S. S. M. Hoque, ``Distributed learning over a wireless network with FSK-based majority vote," \textit{Proc. Int. Conf.Adv. Commun. Technol. Netw.}, Dec. 2021, pp. 1–9.
		
		\bibitem{A5_4} X. Zhao, L. You, R. Cao, Y. Shao, and L. Fu, ``Broadband digital over-the-air computation for asynchronous federated edge learning," \textit{Proc. IEEE Int. Conf. Commun. (ICC)}, May 2022, pp. 5359–5364.
		
		\bibitem{A5_5} A. S¸ahin, B. Everette, and S. S. M. Hoque, ``Over-the-air computation with DFT-spread OFDM for federated edge learning," \textit{Proc. IEEE Wireless Commun. Netw. Conf. (WCNC)}, Apr. 2022, pp. 1886–1891.
		
		\bibitem{A5_6} M. H. Adeli and A. Şahin, ``Multi-cell non-coherent over-the-air computation for federated edge learning," in Proc. \textit{IEEE Int. Conf. Commun.(ICC)}, Seoul, Republic of Korea, May 2022, pp. 4944–4949.
		
		\bibitem{A5_8} A. Şahin and R. Yang, ``Over-the-air computation over balanced numerals," \textit{Proc. IEEE Globecom Workshops}, Dec. 2022, pp. 347–352.
		
		\bibitem{A5_9} M. Tang, S. Cai, and V. K. N. Lau, ``Radix-partition-based over-the-air aggregation and low-complexity state estimation for IoT systems over wireless fading channels," \textit{IEEE Trans. Signal Process.}, vol. 70,pp. 1464–1477, 2022.
		
		\bibitem{A6} S. Razavikia, J. M. Barros da Silva and C. Fischione, ``ChannelComp: A General Method for Computation by Communications," \textit{IEEE Trans. Commun.}, vol. 72, no. 2, pp. 692-706, Feb. 2024.
		
		\bibitem{A7} S. Razavikia, J. M. Barros da Silva and C. Fischione, ``SumComp: Coding for Digital Over-the-Air Computation via the Ring of Integers," \textit{IEEE Trans. Commun.}, Early Access, Aug. 2024.
		
		\bibitem{A12} H. Sifaou and G. Y. Li, ``Over-The-Air Federated Learning Over Scalable Cell-Free Massive MIMO," \textit{IEEE Trans. Wirel. Commun.}, vol. 23, no. 5, pp. 4214-4227, May 2024.
		
		\bibitem{A13} C. Chen, E. Björnson and C. Fischione, ``Over-the-Air Computation in Cell-Free Massive MIMO Systems," \textit{arXiv}, 2024. Available at: https://arxiv.org/abs/2409.00517.
		
		\bibitem{B2} E. Björnson and L. Sanguinetti, ``Making Cell-Free Massive MIMO Competitive With MMSE Processing and Centralized Implementation," \textit{IEEE Trans. Wirel. Commun.}, vol. 19, no. 1, pp. 77-90, Jan. 2020.
		
		\bibitem{A8} H. A. Ammar, R. Adve, S. Shahbazpanahi, G. Boudreau and K. V. Srinivas, ``User-Centric Cell-Free Massive MIMO Networks: A Survey of Opportunities, Challenges and Solutions," \textit{IEEE Commun. Surv. Tutor.}, vol. 24, no. 1, pp. 611-652, Firstquarter 2022.
		
		\bibitem{A9} H. Q. Ngo, A. Ashikhmin, H. Yang, E. G. Larsson and T. L. Marzetta, ``Cell-Free Massive MIMO Versus Small Cells," \textit{IEEE Trans. Wirel. Commun.}, vol. 16, no. 3, pp. 1834-1850, Mar. 2017.
		
		\bibitem{A10} H. Q. Ngo, L. N. Tran, T. Q. Duong, M. Matthaiou and E. G. Larsson, ``On the Total Energy Efficiency of Cell-Free Massive MIMO," \textit{IEEE trans. green commun. netw.}, vol. 2, no. 1, pp. 25-39, Mar. 2018.
		
		\bibitem{A11} E. Björnson and L. Sanguinetti, ``Scalable Cell-Free Massive MIMO Systems," \textit{IEEE Trans. Commun.}, vol. 68, no. 7, pp. 4247-4261, July 2020.
		
		\bibitem{B1} P.~Billingsley, \emph{Probability and Measure}, 3rd~ed. New York, NY, USA: Wiley, 1995.
		
	\end{thebibliography}
\end{document}